\documentclass[a4paper,fleqn,usenatbib]{mnras}

\usepackage{mathptmx}

\usepackage[T1]{fontenc}
\usepackage{ae,aecompl}

\usepackage{graphicx}	
\usepackage{amsmath}	
\usepackage{amssymb}	
\usepackage{upgreek}

\newcommand{\kms}{km\,s$^{-1}$}

\title[3D mapping of the Crab Nebula. I.]{3D mapping of the Crab Nebula
with SITELLE. I. Deconvolution and kinematic reconstruction}

\author[T. Martin et al.]{
T. Martin$^{1,2}$\thanks{E-mail: thomas.martin.1@ulaval.ca},
D. Milisavljevic$^{3}$,
L. Drissen$^{1,2}$
\\
$^{1}$D\'epartement de physique, de g\'enie physique et d'optique,
Universit{\'e} Laval, 2325, rue de l'universit{\'e}, Qu{\'e}bec
(Qu{\'e}bec), G1V 0A6, Canada\\
$^{2}$Centre de Recherche en Astrophysique du Qu{\'e}bec
(CRAQ)\\
$^{3}$Department of Physics and Astronomy, Purdue University,
525 Northwestern Avenue, West Lafayette, IN, USA}

\date{Accepted XXX. Received YYY; in original form ZZZ}

\pubyear{2019}

\DeclareRobustCommand{\NII}{\textup{N\,\textsc{\lowercase{II}}}}
\DeclareRobustCommand{\SII}{\textup{S\,\textsc{\lowercase{II}}}}

\DeclareRobustCommand{\Halpha}{\textup{H$\alpha$}}
\DeclareRobustCommand{\kms}{km\,s$^{-1}$}
\DeclareRobustCommand{\cm1}{cm$^{-1}$}
\DeclareRobustCommand{\degrees}{$^\circ$}

\hypersetup{draft}
\begin{document}
\label{firstpage}
\pagerange{\pageref{firstpage}--\pageref{lastpage}}
\maketitle

\begin{abstract}

We present a hyperspectral cube of the Crab Nebula obtained with the
imaging Fourier transform spectrometer SITELLE on the
Canada-France-Hawaii telescope. We describe our techniques used to
deconvolve the 310\,000 individual spectra ($R = 9\,600$) containing
  \Halpha, [\NII] $\lambda\lambda$6548, 6583, and [\SII]
  $\lambda\lambda$6716, 6731 emission lines and create a detailed
  three-dimensional reconstruction of the supernova remnant assuming
uniform global expansion. We find that the general
boundaries of the 3D volume occupied by the Crab are not strictly
ellipsoidal as commonly assumed, and instead appear to follow a
``heart-shaped'' distribution that is symmetrical about the plane of
the pulsar wind torus. Conspicuous restrictions in the
bulk distribution of gas consistent with constrained expansion
coincide with positions of the dark bays and east-west band of He-rich
filaments, which may be associated with interaction with a
pre-existing circumstellar disk. The distribution of filaments follows
an intricate honeycomb-like arrangement with straight and rounded
boundaries at large and small scales that are anti-correlated with
distance from the center of expansion. The distribution is not unlike
the large-scale rings observed in supernova remnants 3C\,58 and
Cassiopeia\,A, where it has been attributed to turbulent mixing
processes that encouraged outwardly expanding plumes of radioactive
$^{56}$Ni-rich ejecta. These characteristics reflect critical details
of the original supernova of 1054\,CE and its progenitor star, and may
favour a low-energy explosion of an iron-core progenitor. We demonstrate that our main findings are
robust despite regions of non-homologous
expansion driven by acceleration of material by the pulsar
wind nebula.

\end{abstract}




\section{Introduction}

Young supernova (SN) remnants ($< 3000$\,yr) in the Milky Way and
Magellanic Clouds provide rare opportunities to probe the explosion
mechanisms and progenitor systems of SNe with observations capable of
producing three dimensional reconstructions.  Their proximity (from a few to less than a hundred
kpc) and high expansion velocity (a few thousand \kms) allow not only
to disentangle the different Doppler components with relatively modest
spectral resolutions, but also to determine the tangential velocity of
their filaments using images obtained decades apart (or sometimes a
few years apart using the {\sl Hubble Space Telescope}; {\sl
  HST}). Asymmetries in chemical or ionization structure, expansion
velocity or density can then be probed with much greater precision
than what is possible for unresolved extragalactic objects
\citep{MF17}. Such reconstructions, made for SNRs including
Cassiopeia\,A (Cas~A)
\citep{DeLaney10,Mili13,Alarie14,MF15,Grefenstette17}, 1E 0102.2-7219
\citep{Vogt10,Vogt18}, and N132D \citep{Law20}, are critical for
establishing strong empirical links between SNe and SNRs that
can be compared to state-of-the-art simulations evolving from core
collapse to remnant \citep{Orlando15,Orlando16,Orlando20,Ono20}.

Among the most studied yet still enigmatic remnants deserving of 3D
reconstruction is the Crab Nebula (SN 1054, NGC 1952).  Despite decades of
investigation, the progenitor star's initial mass and the properties
governing the SN explosion remain uncertain
\citep{DF85,Hester2008}. The total mass of its ejecta
(2-5\,M$_{\odot}$; \citealt{Fesen1997}) is much less than the
plausible mass of the progenitor (8-13 M$_{\odot}$;
\citealt{Nomoto87}), and although SN\,1054 was more luminous than a
normal Type II SN ($-18$ mag vs.\ $-15.6$ mag; \citealt{CS77}), its kinetic energy ($\approx 7 \times 10^{49}$\,erg) is surprisingly low compared to the
canonical $\sim 10^{51}$\,erg.  The standard explanation is that most
of the mass and 90\% of the kinetic energy of SN 1054 reside in an
invisible freely expanding envelope of cold and neutral ejecta traveling $\sim 5000$ \,km\,s$^{-1}$ far
outside the Crab \citep{Chevalier77}. However, this theorized outer envelope has
never been robustly detected to remarkably low upper limits \citep{Fesen1997,Lund2012}. A weak C~IV $\lambda$1550 absorption feature  observed in a far-ultraviolet {\sl HST} spectrum of the Crab pulsar is suggestive of an ionized outer envelope \citep{Sollerman00,Hester2008}, but only extending out to $\sim 2500$\,km\,s$^{-1}$ and tracing a relatively small amount of material ($\sim 0.3$ M$_{\odot}$).  

Models and
observations generally support a low energy SN origin potentially associated
with an O-Ne-Mg core that collapses and explodes as
electron-capture supernova (ECSN)
\citep{Nomoto82,Hillebrandt82,Kitaura06}. However, such explosions are
generally faint ($M_V$ $> -15$ mag) and thus inconsistent 
with the brightness
of SN 1054 estimated from historical records. \citet{Fesen1997} and \citet{CO00} suggested that SN 1054
was a low energy SN with additional luminosity provided by
circumstellar interaction. \citet{Smith13} supports this view and
identified potential Crab-like analogs in many recent Type IIn
events. \citet{Tominaga13} found that the high peak luminosity could
instead be related to the large extent of the progenitor star and not
necessarily associated with strong circumstellar
interaction. \citet{Gessner18} questioned the appropriateness of an
ECSN origin for the Crab, as their simulations found that hydrodynamic
neutron star kicks associated with O-Ne-Mg core progenitors are much
below the $\sim 160$\,km\,s$^{-1}$ measured for the Crab
pulsar \citep{Kaplan2008}. \citet{YC15} found the Crab's overall
properties to be consistent with expectations from a pulsar wind
nebula evolving inside a freely expanding low energy supernova.

There have been multiple attempts at mapping the three dimensional
structure of the Crab. However, the large angular size of the Crab
($\sim$6\arcmin) and complexity of its numerous overlapping
filamentary structures presents many challenges.
\citet{Lawrence95} created three-dimensional spatial models of the
line-emitting [O~III] $\lambda\lambda$4959, 5007 gas in the Crab with
Fabry-Perot imaging spectroscopy. \citet{Cadez04} created a 3D
representation using long slit spectroscopy at low and high resolution
configurations rotated at a series of position
angles. \citet{Charlebois10} used the imaging Fourier transform
spectrometer SpIOMM to create a 3D view of the Crab and emission line
ratios of the filaments. \citet{BF15} mapped the Crab's northern
ejecta jet with moderate resolution [O~III] line emission
spectra. Generally, these investigations have highlighted the
north-south bipolar asymmetry in the abundance, geometry, and velocity
distribution of the bright filaments. A band of helium-rich material
runs in the east-west direction \citep{UM87}, which is associated with
pinched velocities \citep{MacAlpine89}, potentially associated with
constrained expansion due to interaction with a circumstellar disk
left behind from pre-SN mass loss \citep{Fesen1992}. To date there does
not exist a complete mapping of individual emission lines throughout
the remnant sensitive to faint emission on fine scales.

In this paper we introduce a hyperspectral cube of the Crab obtained
with the imaging Fourier transform spectrometer SITELLE
\citep{Drissen19}. SITELLE has a field of view of 11\arcmin{}$\times$ 11\arcmin{}, high sensitivity down to 350 nm, and is especially powerful for observing emission
line sources above a low continuum background
\citep{Bennett2000,Maillard2013}. Together, these characteristics make
SITELLE uniquely suited to meet the challenges of observing the Crab.

In this paper we describe preliminary SITELLE observations and
associated analysis of the Crab obtained in a passband covering
647-685 nm with spectral resolution 9\,600. These data are the
highest resolution ever obtained with SITELLE on an astrophysical
target and a spectacular opportunity to demonstrate the full potential
of this new technology. We also describe the techniques
used to deconvolve the spectra and produce a 3D reconstruction of
the supernova remnant. A more detailed analysis of these data in
combination with planned complementary observations at other
wavelengths will follow in a subsequent paper.

\begin{figure}
  \includegraphics[width=\linewidth]{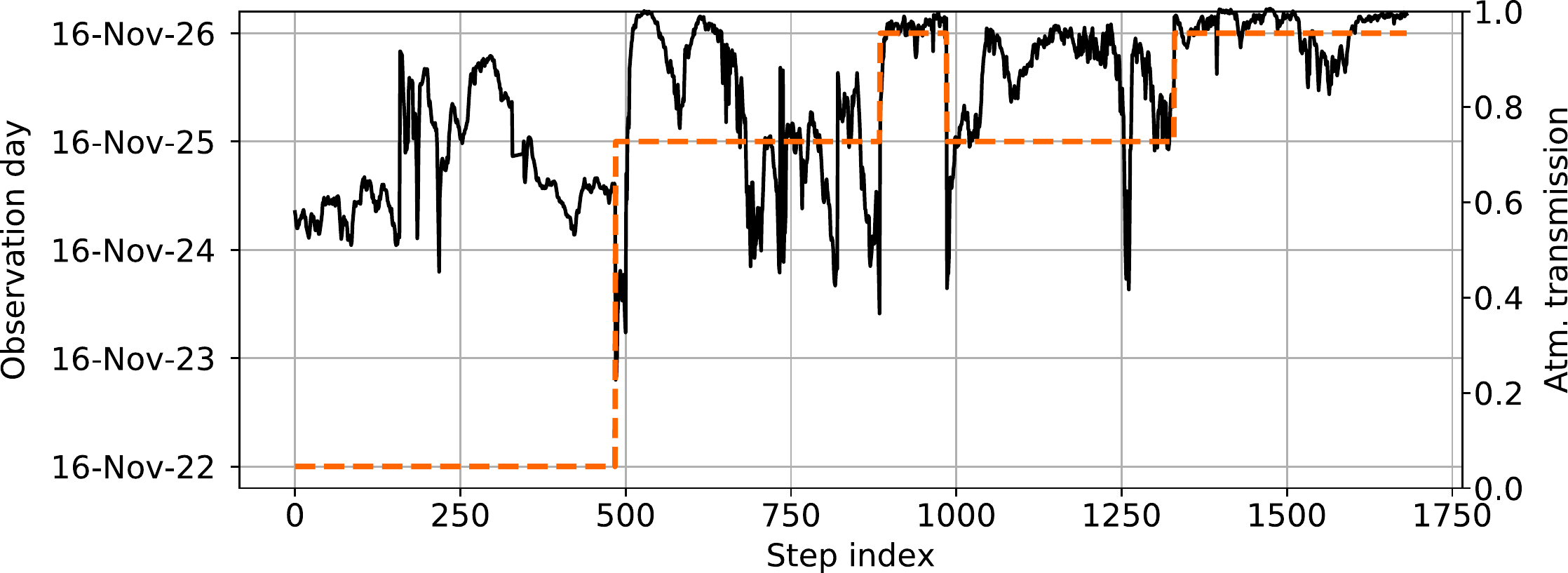}
  \caption{Atmospheric transmission as a function of the step
    index. The date when each part of the cube was obtained is shown
    as an orange dotted line.}
    \label{fig:transmission}
\end{figure}

\begin{figure}
  \includegraphics[width=\linewidth]{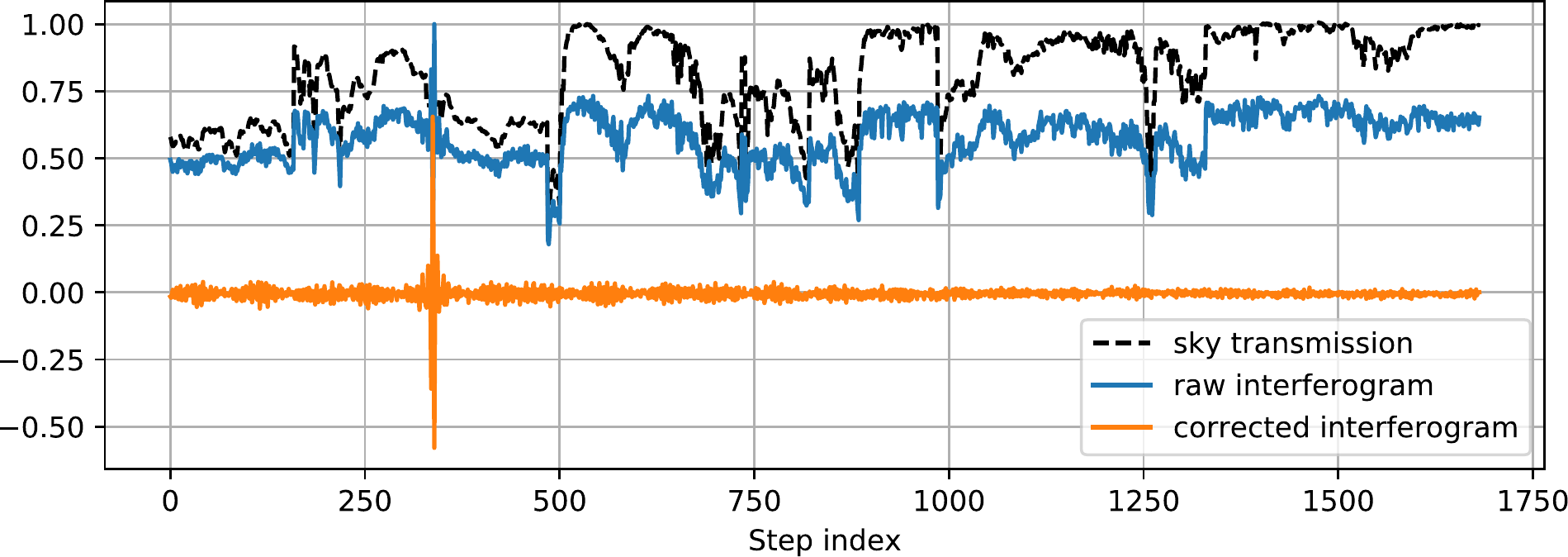}
  \caption{Interferogram of the M1 pulsar before and after the
    correction for the sky transmission and the varying background
    (resp. blue and orange line). Atmospheric transmission is reported
    in dotted black. The large oscillation seen at step $\sim 325$
    corresponds to the Zero path difference of the interferometer, and
    is characteristic of a continuum source, while the low-amplitude
    beating patterns observed along the interferogram are caused by
    the multiple emission lines of the nebula included within the
    $1.6''$ radius of the aperture.}
  \label{fig:pulsarinterf}
\end{figure}

\section{Data}

\subsection{SITELLE data}

The data were obtained using the SITELLE instrument mounted to
Canada-France-Hawaii telescope (CFHT) during the course of an
engineering run spanning the nights of November 22, 25 and 26,
2016. SITELLE combines a 2D imaging detector with a Michelson
interferometer. Two complementary interferometric data cubes are
obtained by recording images, on two 2k$\times$2k CCD detectors, at
different positions of the moving mirror inside the Michelson. Fourier
transforms are then used to convert these cubes into a single spectral
data cube. Spectral resolution is set by the maximum path difference
between the two arms of the interferometer, reached by displacing its
moving mirror through a series of steps of several hundred nanometers
each.  The spectral range is selected by using interference filters;
SITELLE covers the 350 - 850 nm range with a series of 8 filters,
tailored to specific needs.  Spatial sampling is $0.32''$ per pixel,
leading to a field of view of $11' \times 11'$ and over 4 million
spectra.

Our raw data for the Crab consist in an interferometric cube of 1682
steps (with a step size of 2843 nm) with an exposure time of 5.3 s per
step (followed by an overhead of 3.8 s for CCD readout and concurrent
mirror movement and stabilization), leading to an integration time of
2.48 hours over a 4.25 hours total data aquisition time.  The median
seeing, measured on the image obtained from the combination of all
detrended and aligned interferometric images, was
1.17$\pm 0.04$\,\arcsec.  This engineering data was aimed at testing
SITELLE's high resolution capabilities with a target resolution of
10\,000 in the SN3 filter (647 - 685 nm passband: \Halpha{},
[\NII{}]$\lambda\lambda$6548,6584,
[\SII{}]$\lambda\lambda$6717,6731). It was obtained under extremely
varying atmospheric conditions (see
Figure~\ref{fig:transmission}). Some data obtained on November 25
(around steps 900 - 1000) were deemed to be of too poor quality and
were therefore taken again at the end of the following night.

We thus have been able to test the stability of the instrument in
terms of absolute positioning of the moving mirror and the impact of
the observed modulation efficiency loss at high optical path
difference (OPD) \citep{Baril2016}.

These data were reduced with the pipeline reduction software
\texttt{ORBS} \citep{Martin2012,Martin2015-thesis} without any special
treatment with respect to the rest of the data obtained during the
same run. As an example of the quality of the reduction, we present in
Figure~\ref{fig:pulsarinterf} the raw interferogram of the pulsar,
obtained with an aperture of $1.6''$ radius, before and after the
correction for the sky transmission. Any error on this correction on
the interferograms can significantly impact the quality of the calculated
spectrum and especially its instrumental line shape (ILS)
\citep{Martin2016b}.

\begin{figure}
  \includegraphics[width=\linewidth]{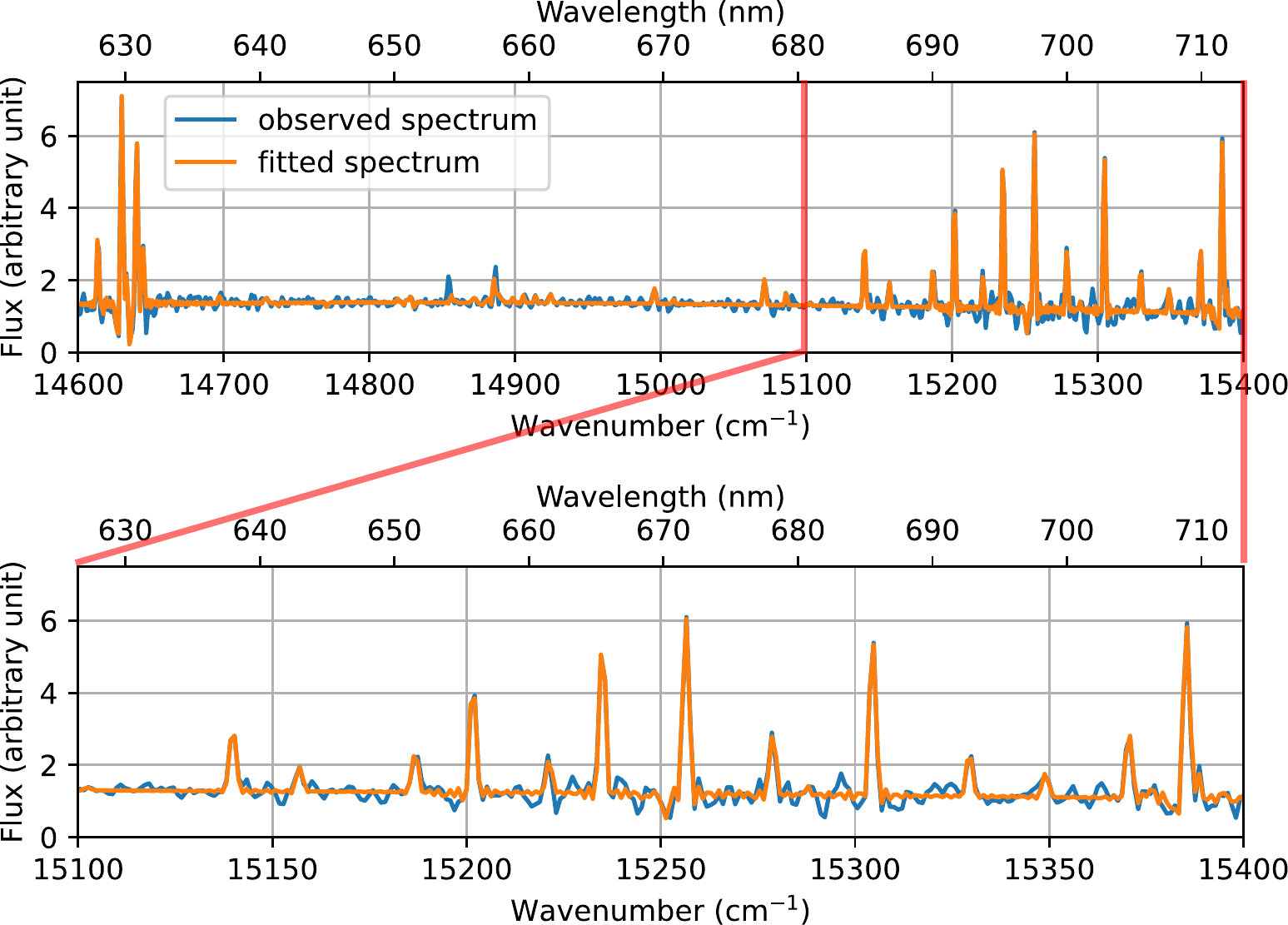}
  \caption{Sky spectrum used to measure the effective resolution
    ($R=8870\pm20$) (blue line) along with the fitted model (orange
    line). The bottom panel shows an enlarged portion of the top
    panel around \Halpha{} (15237\,cm$^{-1}$)}
  \label{fig:M1_sky_spectrum_fit}
\end{figure}

We have measured the effective resolution by fitting a model on a
spectrum of the sky (dominated by OH lines in this spectral range)
integrated over a small region of the cube (a circular aperture with a
100 pixels radius) with the analysis software \texttt{ORCS}
\citep{Martin2015}. In order to take into account the modulation
efficiency loss at high OPD, which may broaden the observed ILS, we
have modelized the ILS as the convolution of a sinc (the natural ILS
of a Fourier transformed spectrum) with a Gaussian (resulting from the
broadening of the observed sky lines by the modulation efficiency
loss) and used the \texttt{sincgauss} model described in
\citet{Martin2016} (see Figure~\ref{fig:M1_sky_spectrum_fit}). The
measured full-width at half maximum of the sky lines was
1.718$\pm0.05$\,\cm1{} which leads to a resolving power $R=8870\pm20$
@ \Halpha{} \citep{Martin2016}. From the parameters of the cube, the
theoretical resolution which we should have been measured at the same
position is 9640. We can conclude that the effect of the modulation
efficiency loss at high OPD is at most of the order of 10\,\% at a
resolution of 9640.

\section{Mapping the Crab Nebula in \Halpha, [N~II] and [S~II]}

\begin{figure}
  \includegraphics[width=\linewidth]{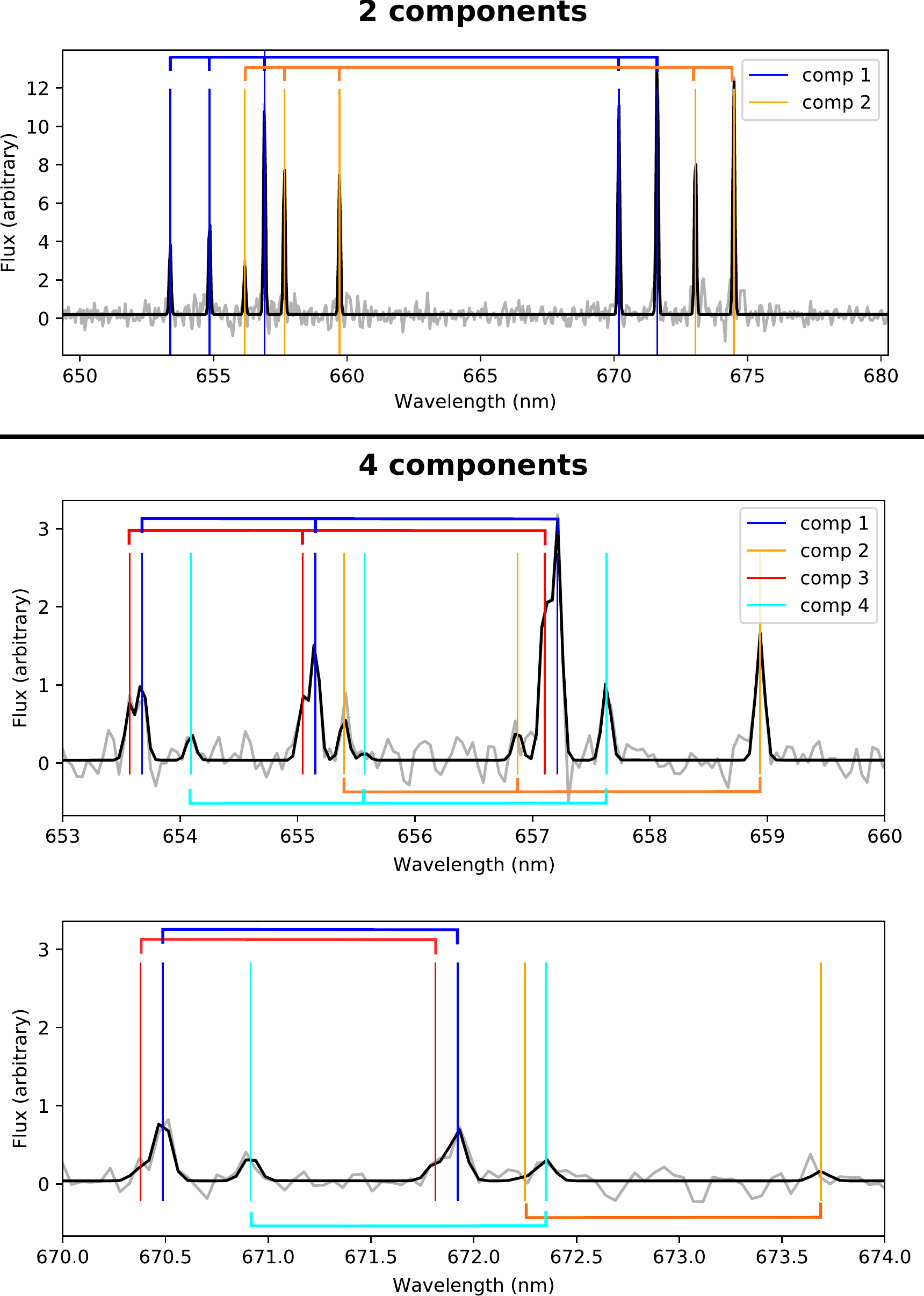}
  \caption{Two examples of spectra containing multiple components and
    their fit. \textbf{2 components panel: } A spectrum with 2 non-overlapping
    components, obtained near the center of the
    nebula. \textbf{4 components panel: } A spectrum with 4 overlapping components
    obtained in a complex filamentary region of the nebula. The plot is
    split in two parts to help distinguish all 4 components. Note
    that if only 4 of the components have been found by the algorithm,
    additional dim components may exist. This example demonstrates at
    the same time the quality and the limitations of our algorithm.}
  \label{fig:all_comps}
\end{figure}

The Crab Nebula is long known to display a very complex filamentary
structure that remains interpreted as the result of Rayleigh-Taylor
instabilities at the interface of the synchrotron nebula and the
thermal ejecta \citep{Hester1996,Hester2008}. This filamentary
structure, ionized by the shock of the expanding synchrotron nebula,
shows's particularly strong emission in \Halpha{}, [\NII{}] and
[\SII{}]. Multiple components of filamentary emission are visible
along any line of sight with velocities ranging from -1500 to
1500\,\kms{} (see Figure~\ref{fig:all_comps}) that must be separated
in order to compute the correct mapping of the flux and velocity of
the observed emission lines. The emission covers a circular surface of
$\sim$6\,\arcmin{} in diameter, which spans approximately one million
of the four million spectra contained in the data cube.

\begin{figure*}
  \includegraphics[width=\linewidth]{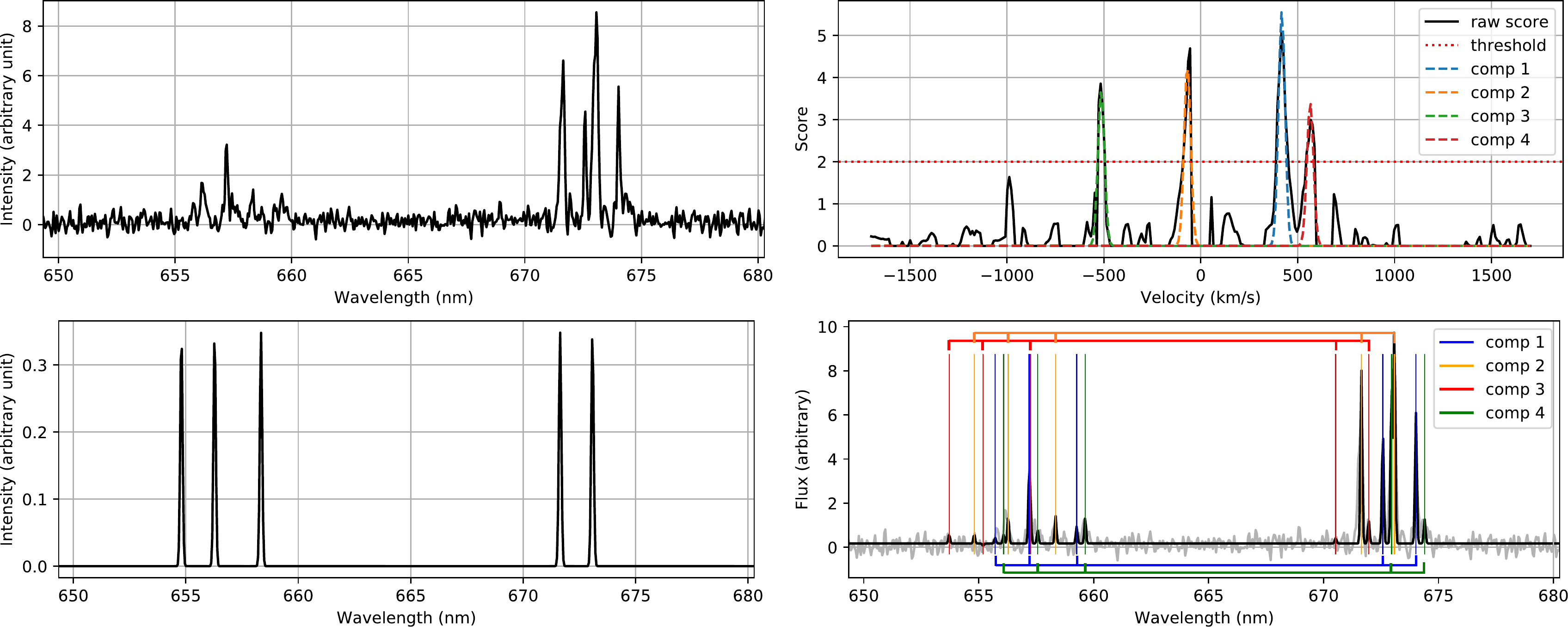}
  \caption{The fitting process. \textit{Top-left:} Example of a raw spectrum
    extracted from a complex filamentary region of the nebula containing at least 4 components. \textit{Bottom-left:} Example of a 5
    emission-lines comb-like spectrum which would be convolved with
    the analyzed spectrum. In fact, the spectrum is convolved with the
    individual lines of the comb. All lines of the comb have the same
    intensity. The apparent difference comes from the function
    sampling. \textit{Top-right:} Mitigated correlation score obtained
    after the first step of the analysis (black) with the 4 fitted
    components detected at the second step. The detection threshold used for the enumeration of the brightest components of the score (see section~\ref{sec:step2}) is
    shown in dotted red. \textit{Bottom-right:} Fit realized on the
    spectrum shown in the top-left panel; the positions of the 5
    fitted emission-lines are shown for each of the 4 fitted
    components.}
  \label{fig:all_comps_2}
\end{figure*}

Inspired by the algorithms developed by \citet{Cadez04} and \citet{Charlebois10} on similar data sets, we have written an algorithm to automate detection and fitting of overlapping emission components observed in the spectra. This algorithm analyses each spectrum individually in 3 steps (see Figure~\ref{fig:all_comps_2}):
\begin{enumerate}
\item evaluation of the probability, as a score, of having one
  component at a given velocity;
\item enumeration of all the individual velocity components along the
  line of sight;
\item fit of the spectrum with a model combining all the velocity
  components at the same time.
\end{enumerate}

\subsection{Step 1: computation of the score}

The first step is based on the convolution of the analyzed spectrum
$S(\lambda)$ with a comb-like spectrum $K(v, \lambda)$ made of a
subset of the emitting lines of each component: \Halpha{}, the
[\NII{}] doublet and the [\SII{}] doublet (see bottom-left panel of
Figure~\ref{fig:all_comps_2}). All lines of the comb have the same
amplitude. The calculated score $C(v)$ is simply:
\begin{equation}
  \label{eq:basic}
  C(v) = \int S(\lambda) K(v, \lambda) \text{d}\lambda\;.
\end{equation}
$C(v)$ is maximum when the position of the modeled emission-lines of
$K(v, \lambda)$ coincides with the position of the emission-lines of
the spectrum. 

Ideally, if the explored velocity range is not too large and no line of the comb is matched with another emission-line, each
velocity component of the spectrum will produce one peak with an
approximately Gaussian shape. The centroid of the peak gives the
component velocity and its amplitude scales with the integral of the
flux in the lines present in $K$.

However, the biggest challenge with this approach appears when the comb and
the analyzed spectrum contains multiple lines within the range of
velocity scanned. For example, one line
of the comb (e.g., \Halpha{}) may coincide with the position of a
neighbouring line at a different velocity (e.g., [\NII{}]). In this case, even with only one
component along the line of sight, $C(v)$ will show multiple peaks; the highest being the real one
because it reflects the velocity at which the largest number of lines
are coincident. With multiple components however, if one component is
much brighter than the others, the secondary peaks in $C(v)$ created
by the brightest component may be even higher than the primary peak of
the second components, in which case the correct enumeration of the
components is compromised.

\begin{figure}
  \includegraphics[width=\linewidth]{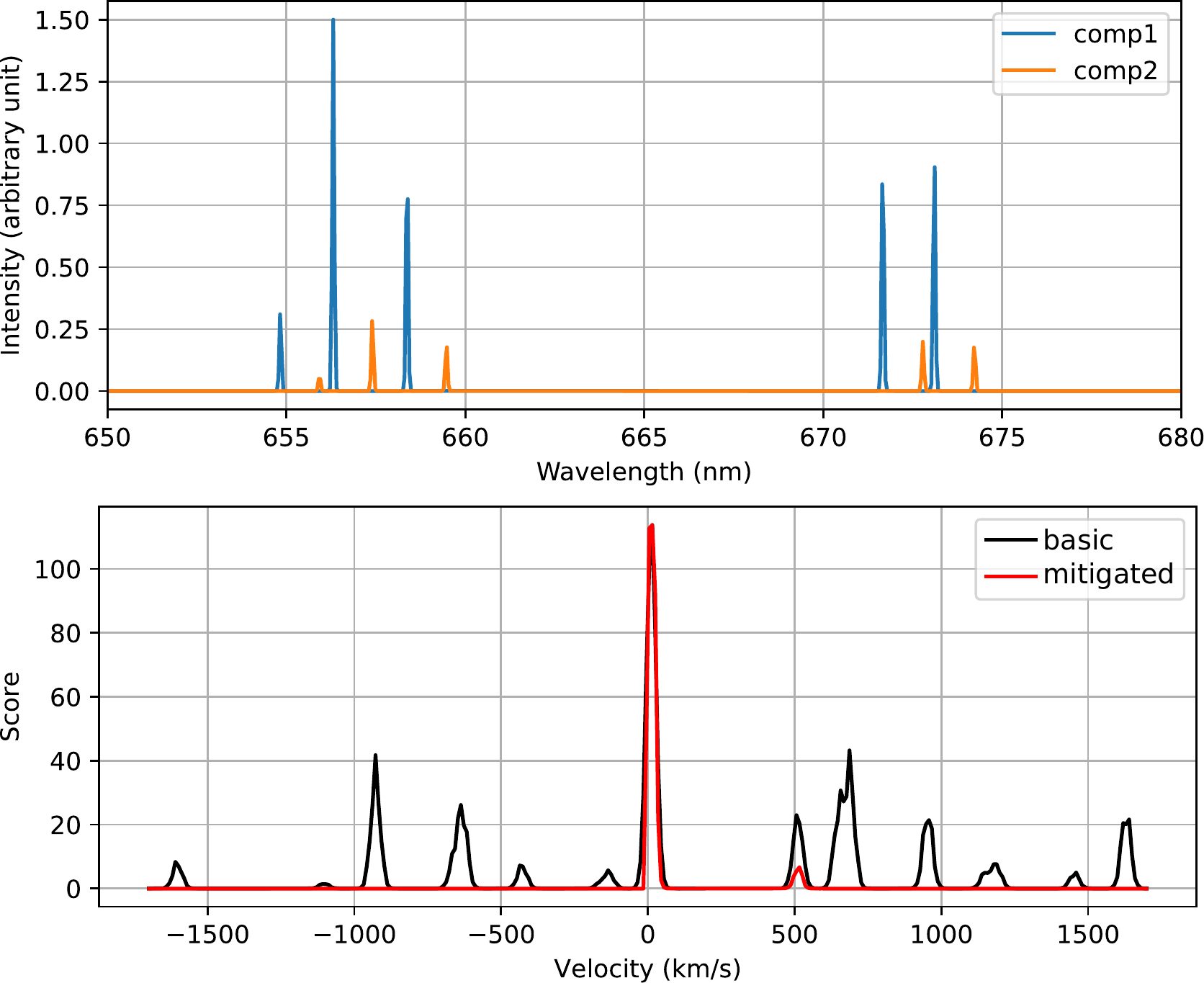}
  \caption{\textit{Top:} Two-components noiseless model
    spectrum. \textit{Bottom:} Basic and mitigated scores computed from
    the top-panel spectrum. The numerous lines appearing in the basic score (black line) do not reflect real components or noise. Instead, they correspond to velocities where some lines of the comb coincide with another emission line (e.g., where the \Halpha{} line of the comb coincides, at a given velocity, with an [\NII{}] line of the spectrum). Applying physical constraints when computing a mitigated score (in red) reduces these aliases, which permits the detection of dimmer velocity components. The addition of noise only contributes  small amplitude peaks that are discarded (see top-right panel of Figure~\ref{fig:all_comps_2}). }
  \label{fig:model_example}
\end{figure}

Figure~\ref{fig:model_example} reproduces this issue by showing a
synthetic spectrum made of two velocity components, one being 5 times
brighter than the other. The comb used for the analysis contains 5
emission lines (\Halpha{}, [\NII{}], [\SII{}]) and is shown in the
bottom-left quadrant of Figure~\ref{fig:all_comps_2}.

Using equation~\ref{eq:basic} without any special treatment leads to a
score $C(v)$ with multiple false peaks (black line) where the peak
related to the dimmest component cannot be retrieved.

We mitigate complicating factors with use of equation~\ref{eq:basic}
by adding a number of physically-based conditions that must be
respected in order to get a non-zero value of $C(v)$. One is to force
the presence of all the lines by computing independent scores for each
line and compute the product of their probability. Let
$K_i(v, \lambda)$ be the kernel for the line $i$, $i \in \{\Halpha,$
$[\NII]\lambda 6548,$ $[\NII]\lambda 6548,$ $[\NII]\lambda 6583,$
$[\SII]\lambda 6716,$ $[\SII]\lambda 6731\}$. If we want all 5 lines
to be present in order to have a non-zero score we would rewrite
equation~\ref{eq:basic} as
\begin{gather}
  \label{eq:product}
  C_i(v) = \int S(\lambda) K_i(v, \lambda) \text{d}\lambda\;,\\
  C(v) = \Pi_{i}\; C_i(v)\;.
\end{gather}

\begin{figure*}
  \includegraphics[width=.75\linewidth]{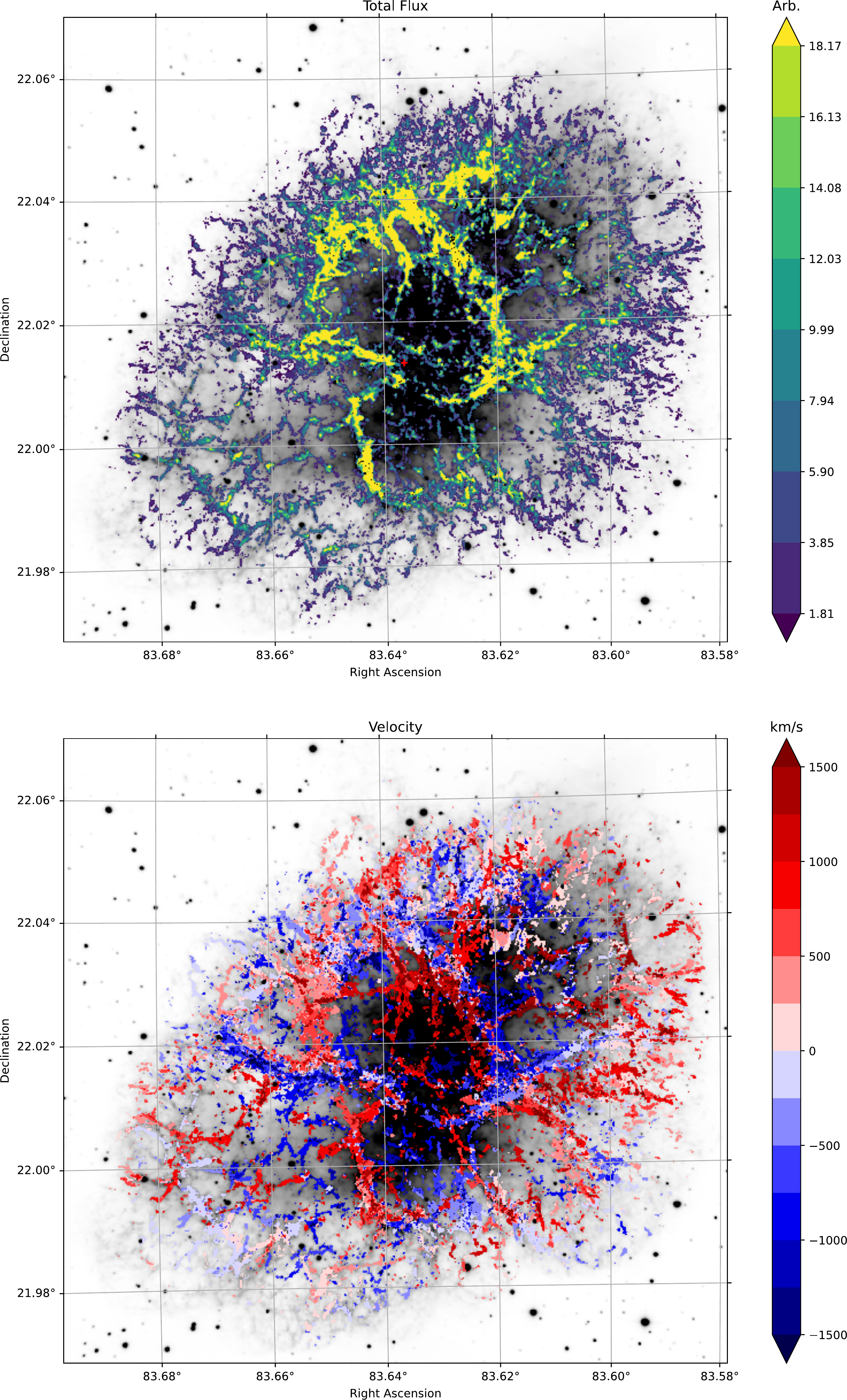}
 
  \caption{Integral of the emission in all five emission lines (\textit{top}) and velocity (\textit{bottom}) maps obtained. When
    components are overlapping, only the front component is
    displayed. The color mapping has been discretized for the sake of
    clarity, but it does not reflect the actual precision of the
    values. The red cross indicates the center of expansion computed by \citet{Kaplan2008}. The velocity uncertainty is shown in Figure~\ref{fig:fig_all_err}. The background is the integral of the flux measured (emission lines + continuum) in the whole filter.}

\label{fig:fig_all}
\end{figure*}

However, all 5 lines are not always detectable. Sometimes only the
$[\SII]$ or $[\NII]+{\Halpha}$ lines are visible. Thus we further separate
these two groups and put additional physics-based conditions to
finally write the mitigated version of the score:
\begin{gather}
  \label{eq:mitigated}
  C_{[\NII]} = \sqrt{C_{[\NII]\lambda 6548} \times C_{[\NII]\lambda 6583}}\;,\\
  C_{[\SII]} = \sqrt{C_{[\SII]\lambda 6716} \times C_{[\SII]\lambda 6731}}\;,\\
  C(v) = C_{\Halpha} \times C_{[\NII]} + C_{[\SII]}\;;
\end{gather}
with the following additional conditions that constrain the [\SII{}]
and [\NII{}] line ratios to be realistic enough,
\begin{gather}
  C_{[\NII]} = 0\text{ if }C_{[\NII]\lambda 6583} < C_{[\NII]\lambda 6548}\;,\\
  C_{[\NII]} = 0\text{ if }C_{[\NII]\lambda 6548} < 0.14 \times C_{[\NII]\lambda 6584}\;,\\
  C_{[\SII]} = 0\text{ if }C_{[\SII]\lambda 6716} < 0.14 \times C_{[\SII]\lambda 6731}\;,\\
  C_{[\SII]} = 0\text{ if }C_{[\SII]\lambda 6731} < 0.14 \times
  C_{[\SII]\lambda 6716}\;.
\end{gather}
The value of 0.14 ($\simeq 1/7$) has been manually optimized to help reject obviously
wrong scores while keeping lower SNR components. This value is necessarily kept lower than the theoretical ratios of [\SII] lines (between 0.5 and 1.5) and [\NII] lines (2.94 in the low-density regime) \citep{OsterbrockDonaldE.2006} to take into account noise associated with the measured flux.

The results of this mitigated score is drawn on
Figure~\ref{fig:model_example} with a red line. The real peaks are
both present and all the false peaks have been removed. A more
realistic example of this score is also shown in the top-right
quadrant of Figure~\ref{fig:all_comps_2}, where the false peaks, if
not completely removed, have been attenuated enough so that the 4
brightest components are clearly visible.

\subsection{Step 2: enumeration of the brightest components}
\label{sec:step2}

Once the score is obtained, we must go through a peak detection
process to enumerate the brightest components and evaluate their
velocity, which is done by measuring the centroid of each detected
peak. As the components may have similar velocities, their peaks may
overlap, which complicates the detection.

We have used an iterative detection procedure not unlike the CLEAN algorithm \citep{hogbom_aperture_1974}. At each iteration, only
the brightest peak is detected, fitted and removed before moving on to
the next iteration until no peaks can be detected above a
threshold. The threshold was manually adjusted to keep the number of
false detections negligible at the expense of loosing some of the
dimmest components.  An example of the resulting detection is shown in
the top-right quadrant of Figure~\ref{fig:all_comps_2}, where only 4
of the possibly 5 components are bright enough to be considered. Since the noise of an FTS spectrum is distributed over all the channels and proportional to the total flux of the source, using a variable threshold based on our knowledge of the noise level for each spectrum may help in detecting components in dimmer regions of the Crab. This possibility will be explored in future versions of our algorithm.

At the end of this step we detected emission in 310\,000 pixels (of
the $\sim$1~million spectra analyzed). 73.2\,\% of them have only one
component along the line of sight, 20.7\,\% contain 2 components,
4.95\,\% contain 3 components and less than 1 percent contain more
than 3 components.

\subsection{Step 3: fit of the spectrum}
\label{sec:step3}
Once all the components are enumerated, a fit of the whole spectrum is
attempted. This fit is done with \texttt{ORCS}, a Python module designed especially to fit the spectra obtained with SITELLE
\citep{Martin2015}. Given that the effective resolution is 10\,\%
smaller than the theoretical resolution, the secondary lobes of the
sinc ILS are small enough that a simple Gaussian model can be
used. Five emission-lines are fitted for each component: \Halpha, the
[\NII] $\lambda\lambda$6548, 6583 doublet and the [\SII]
$\lambda\lambda$6716, 6731 doublet. Emission-lines are fitted with a complete spectrum model at a fixed velocity for each component. The FWHM is fixed at the measured effective
resolution. The flux ratio between the [\NII] lines is also fixed at
3. Consequently, only 5 parameters are fitted for each component: 1 for the amplitude of the [\NII] doublet, 3 for the amplitudes of the other lines, and 1 for the velocity of all 5 lines. An example of the resulting fit is shown in the bottom-right
quadrant of Figure~\ref{fig:all_comps_2}. We can see that most lines
are well-fitted except for a possible 5$^{\text{th}}$ component which
was neglected at the enumeration step because the threshold has been
kept high enough to minimize the risks of false detections. 

The data obtained after the automatic fitting procedure is a set of 25
flux maps (5 emission lines $\times$ 5 components, as no more
than 5 components were clearly detected along the line of sight and
only in a few cases) and 5 velocity maps (one for each component,
since each set of emission lines were considered to share the same
velocity parameter). Figure~\ref{fig:fig_all} shows the velocity
mapping and the relative total emitted flux in all the lines. All the components have been combined in the same image
but, for the sake of clarity, when components are overlapping, only
the front component is shown.

\subsection{Step 4: Mapping the Crab Nebula in Euclidean space}

The remarkable work of \citet{Trimble1968} demonstrated that proper
motion velocity vectors of filaments all share the same origin both in
position and time, and that the expansion velocity is approximately
proportional to the radius.  Subsequent analyses have come to similar
conclusions \citep{Wyckoff1977,Nugent1998,Kaplan2008,Bietenholz2015},
though their results on the location of the explosion center or the
mean expansion velocity differ by a few arcseconds \citep{Kaplan2008}
(see Table~\ref{tab:comparison}). The expansion model we adopt is
based on 3 parameters: the right ascension and declination
$(\alpha_c,\delta_c)$ of the expansion center and the expansion factor
$e$ (e.g., \citealt{Bietenholz2015}):
\begin{gather}
  \label{eq:expansion_model}
  \mu_{\alpha} = e (\alpha - \alpha_{c})\;,\\
  \mu_{\delta} = e (\delta - \delta_{c})\;,
\end{gather}
where $\alpha$ and $\delta$ are the coordinates of the filament and
$\mu_{\alpha}$, $\mu_{\delta}$ denote the proper motion along the right
ascension and declination axes. 

It has long been known that when the measured expansion velocity is
projected back to the origin, the computed outburst date lies around
1130\,CE, which is nearly a hundred years after the recorded outburst
date of 1054\,CE \citep{Stephenson2002}. This is attributed to material
having been accelerated by the Crab's pulsar wind nebula.  Thus, we
can expect some sort of signature of this acceleration preferentially
near the center of the explosion. From the preliminary results of a
new analysis of the proper motion of the Crab (Martin et al., in
preparation), we believe that there indeed might be an accelerated
expansion near the center, which means that the expansion factor is
higher near the center than it would be if following a purely linear
model. However, as a first approximation we choose to consider a
simple linear model and use the expansion factor
$e=1.160(15)\times10^{-3}$\,year$^{-1}$ computed by \citet{Nugent1998}
along with the expansion center
$\alpha=05^\text{h}34^\text{m}32.74(03)^\text{s}$,
$\delta=+22^\circ00\arcmin{}47.9(0.4)\arcsec{}$ determined by
\citet{Kaplan2008} from their study of the pulsar proper motion.
The validity of this hypothesis is discussed in more detail in
section~\ref{sec:outer_envelope}.

\begin{table}
  \center

 \label{tab:comparison}
 \begin{tabular}{lccc}
   \hline
   \hline
   Reference& $\Delta\alpha$ & $\Delta\delta$ & Outburst\\
            &(arcsec)&(arcsec)&Date (CE)\\
   \hline
   \citet{Trimble1968}&7.6(1.7)&-8.5(1.4)&1140(15)\\
   \citet{Wyckoff1977}&8.2(2.7)&-8.6(3.6)&1120(7)\\
   \citet{Nugent1998}&9.4(1.7)&-8.0(1.3)&1130(16)\\
   \citet{Kaplan2008}&8.4(0.4)&-8.1(0.4)&\\
   \hline
 \end{tabular}
 \caption{Comparison of the expansion parameters computed by several
   authors. Most of the data comes from Table~4 of
   \citet{Nugent1998} and Table~3 of \citet{Kaplan2008}. Following
   \citet{Kaplan2008}, $\Delta\alpha$ and $\Delta\delta$ are the offset
   in right ascension and declination between the center of the
   explosion and the star 5\arcsec to the northeast of the pulsar
   (which was first used by \citealt{Trimble1968} as a reference
   given its proximity to the center and its small proper
   motion). The coordinates of the reference star
   $\alpha=05^\text{h}34^\text{m}32.1827^\text{s}$,
   $\delta=+22^\circ00\arcmin{}56.002\arcsec{}$ come from the GAIA
   DR2 catalog \citep{Brown2018}.}
\end{table}

\begin{figure*}
  \includegraphics[width=\linewidth]{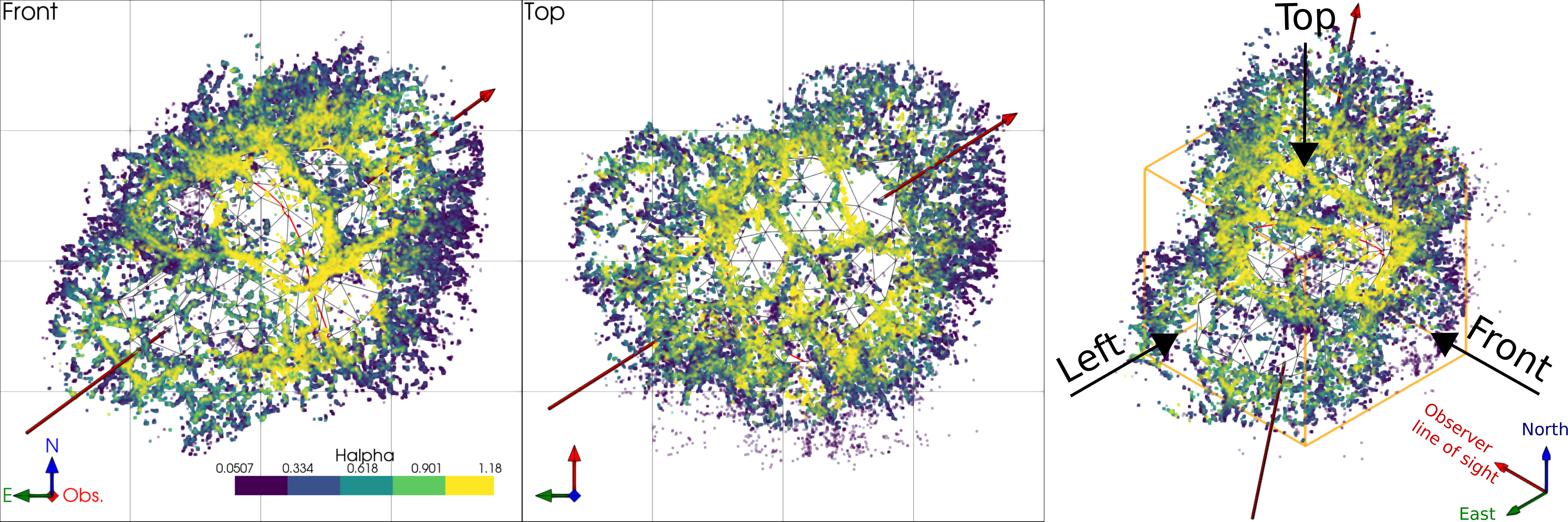}
  \caption{Front (as seen from Earth) and top view of the \Halpha{}
    emission. The right panel displays an isometric representation of
    the nebula showing the different viewpoints considered in the
    paper.  The central part is filled with the inner envelope (see
    section~\ref{sec:inner_envelope}) so that the rear-facing components
    are obscured. The red arrow represents the pulsar torus axis and
    the red line shows the intersection of the inner surface with the
    pulsar torus plane as fitted by \citet{Ng2004}. All axes are in
    parsecs and the spatial grid of the left and central panels has a 1 pc stepping. The orientation symbol is color coded: East is green, North is blue and the line-of-sight direction is red.}
  \label{fig:all_3d}
\end{figure*}

To construct a 3D mapping of the Crab Nebula we require an estimate of
its distance. We adopt 2\,kpc, computed by \citet{Trimble1973}, who
estimates it to lie between 1.7\,kpc and 2.4\,kpc from prior
morphological considerations. This distance has not been improved
since then and is still in use in recent articles (see
e.g., \citealt{Hester2008,Kaplan2008}). At this distance
1\,\arcsec{}=9.696$\times 10^{-3}$\,pc and the radial distance $d$ to
the expansion center can be computed from the radial velocity $v_r$ via the
expansion factor \citep{Ng2004}:
\begin{equation}
  d  = \frac{v_r}{e}
\end{equation}

Knowing the expansion factor and the distance to the Crab makes it
possible to obtain a mapping of our data in the Euclidean space (in
parsecs) as shown in Figure~\ref{fig:all_3d} and
appendix~\ref{sec:3dmaps}. Given the complexity of our data we have
created an interactive visualization in \texttt{Python} accessible
through a Jupyter Notebook. It may be found at
\url{https://github.com/thomasorb/M1_paper}. The 3D visualization
program can also be run directly in any html browser at
\url{https://mybinder.org/v2/gh/thomasorb/M1_paper/master} and does
not require any particular computing knowledge.

Two movies have been created with the help of \texttt{panda3d}, an
open-source framework for 3D rendering \citep{Goslin2004}. They are available online as supplementary material. Both
show the total flux emitted in all 5 emission lines. Each data point
is represented as a small cube and corresponds to one velocity
component at a given pixel. No data points have been added by
interpolation. The thickness of some of the brightest filaments along
the line of sight comes from the fact that they could be resolved and
fitted with two components instead of one. The Milky Way background has been
adequately positioned to simulate what would be the typical perspective
of someone moving around the nebula. The Milky Way map has been
created by the NASA/Goddard Space Flight Center Scientific
Visualization Studio (\url{https://svs.gsfc.nasa.gov/4851}) based in
part on the data obtained with Gaia \citep{Brown2018}. One of the
movies shows a glowing sphere at the center
to simulate the blue continuum emitted by the pulsar wind nebula with an intensity to roughly match that observed in the composite {\sl HST} image presented by
\citet{Loll2013}. The soundtrack is a sonification of the data set. Using the interferograms directly as a sound wave, we have mixed multiple samples played at different rates. The volume of the samples is related to the square of the distance to the nebula and the playing speed is related to the velocity of the observer with respect to the nebula. A second movie highlights the geometry of
the Crab Nebula. The obtained data is shown with the inner and outer
envelopes described in the next section. The pulsar axis and the plane
of the pulsar torus are indicated.

\section{Morphology of the Crab Nebula}

\subsection{Inner and outer envelopes}

\subsubsection{Outer envelope}
\label{sec:outer_envelope}

Each of the \Halpha{}, [\NII]$\lambda\lambda$ 6548, 6584, and
[\SII]$\lambda\lambda$6716, 6731 3D maps we have created is a
collection of volume elements (voxels) that occupy the same volume in
space and for which we can measure an emission line flux. This volume
is bounded by the surface of 1 pixel at the distance of the nebula
(0.32\,\arcsec{} at 2\,kpc i.e., $3.1\times 10^{-3}$\,pc) and an element of
spectral resolution along the line of sight (35\,\kms{}
i.e. $3.1\times 10^{-2}$\,pc), which yields a voxel volume of
$3.0\times 10^{-7}$ \,pc$^3$.

\begin{figure}
  \includegraphics[width=\linewidth]{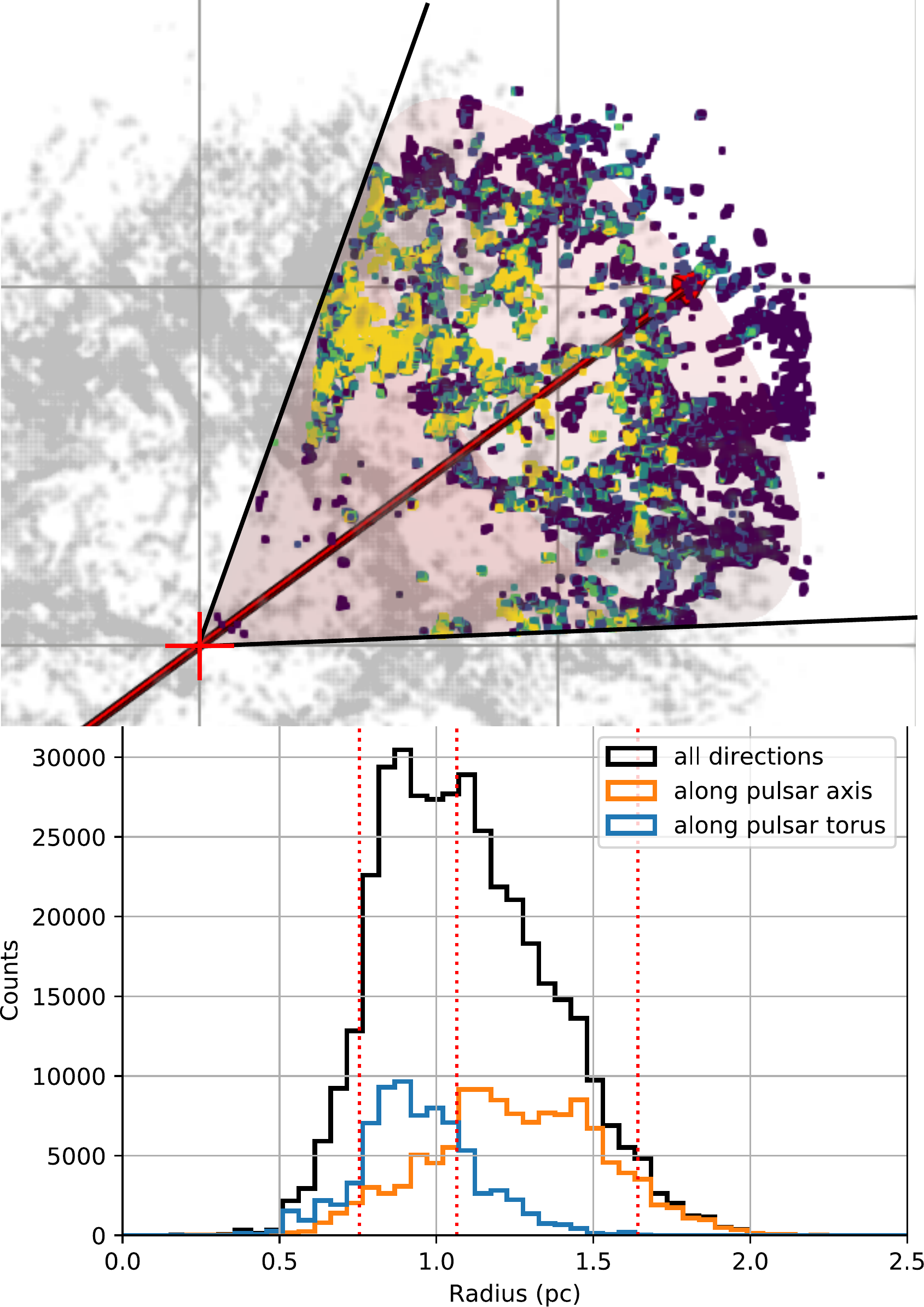}

  \caption{Representation of the emitting material contained in a
    solid angle originating from the explosion center and aligned
    along the pulsar axis. All axes are in parsecs and the spatial
    grid has a 1 pc stepping.}

  \label{fig:cone_example}
\end{figure}

Representing our observations as voxels in this way provides a detailed representation of the Crab's complex distribution of material and permits an investigation of its morphology at small and large scales. For example, it is worth testing whether or not the Crab is an ellipsoid, as long suspected
(e.g., \citealt{Hester2008}). To this end we can analyse the radial distribution of the
emitting material contained in a solid angle originating from the
explosion center and obtain the radial extent of the nebula by
computing the outer limit of this distribution in all directions. To illustrate, we show in Figure~\ref{fig:cone_example} the distribution
of material integrated over all directions along the pulsar torus
axis. We see that no material extends beyond 2\,pc.

Because the emission $I(\Halpha)$ is very roughly proportional to the
square of the ionized hydrogen density $n_p$ (at a constant
temperature along the line of sight), $I(\Halpha) \propto n_p^2$, we
choose to weight this distribution by $\sqrt{I(\Halpha)}$ in order to
approximately sample the distribution of the material density.

\begin{figure}
  \includegraphics[width=\linewidth]{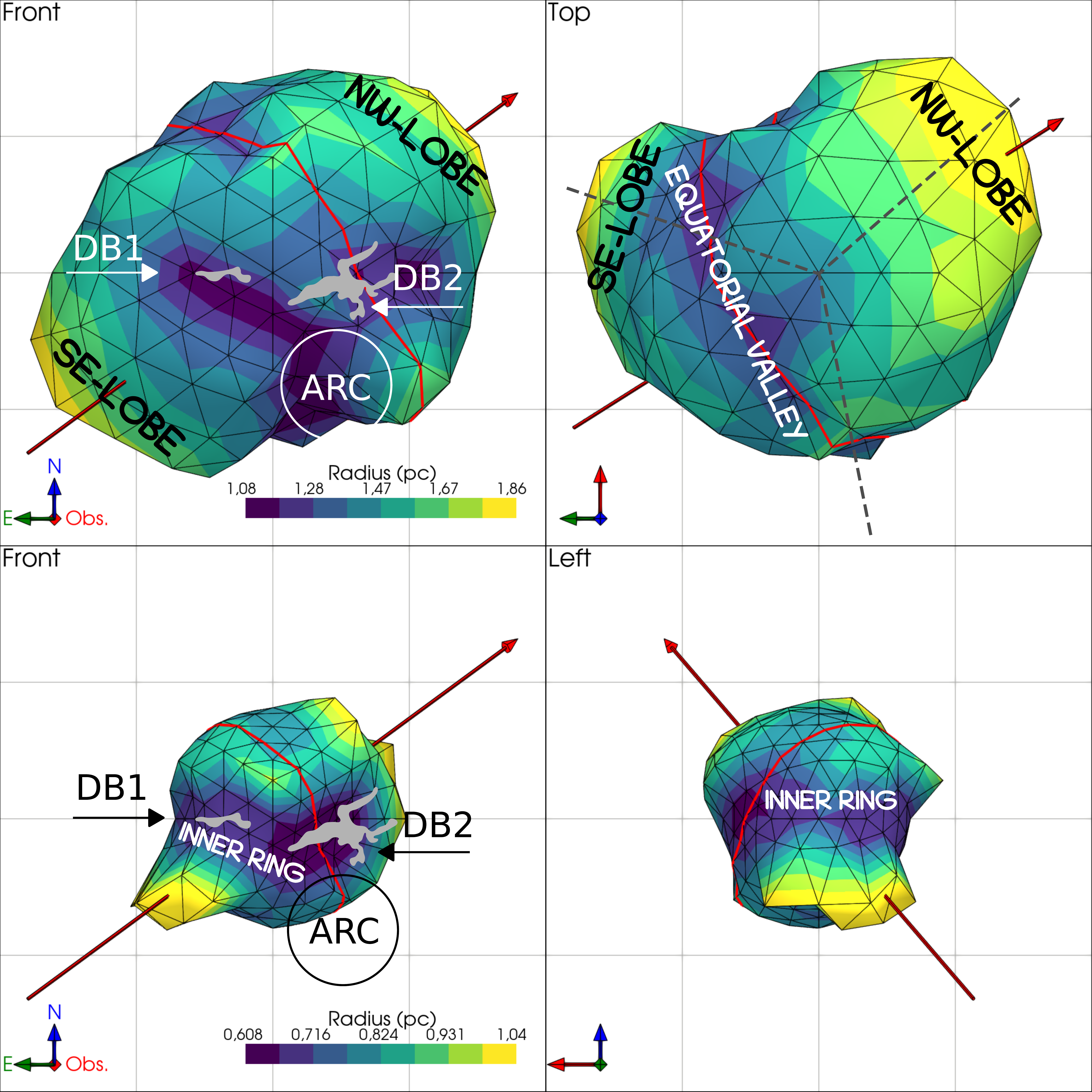}
  \caption{\textit{Top:} Outer envelope enclosing 97\% of the material
    emitting in \Halpha{}. The binning angle covers
    22.5\degrees. \textit{Bottom:} Inner envelope enclosing 8\% of the
    material emitting in \Halpha{}. The binning angle covers
    45\degrees. The positions of the dark bays \citep{Fesen1992} and
    the arcade region \citep{Dubner2017} are shown as well as the
    High-Helium bands first observed by \citet{Uomoto1987} (in gray). We have
    also reported the topologic features referenced in the text.
    Other details are the same as Figure~\ref{fig:all_3d}.}
  \label{fig:innerouter}
\end{figure}

We define the outer limit of the nebula in one particular direction as
the radius that encompasses 97\% of the material emitting in \Halpha{}
which is contained in a solid angle of 22.5\degrees. We repeat
this procedure over all directions, using the coordinates of the
vertices of a 320-faced icosphere which ensures an homogeneous
distribution of the probing directions (see
appendix~\ref{fig:icosphere} for more details). Consequently, we
obtain the outer envelope of the emitting material, i.e., the dominant
shape of the Crab as made up by its visible gas. Two perspectives of
this outer limit surface are shown in Figure~\ref{fig:innerouter}, and
all six perspectives are shown in Figure~\ref{fig:outer6}.

Note that, doing this in the plane of the sky and considering only the red
part of the visible spectrum would reveal the well known elliptical outer
envelope of the Crab. But the 3D outer envelope is most surprising since it differs notably from
the generally assumed ellipsoidal shape
(e.g., \citealt{Hester2008}). As viewed from the {\it top}, which we
define as being along the axis perpendicular to the observer's line of
sight looking from the north toward the south, a conspicuous
heart-shaped morphology oriented along the pulsar axis is visible. The
most rapidly expanding NW and SE lobes are separated by 120\degrees{}
of each other. The NW lobe is nearly aligned with the pulsar torus
axis, but the SE lobe is not.

\begin{figure}
  \includegraphics[width=\linewidth]{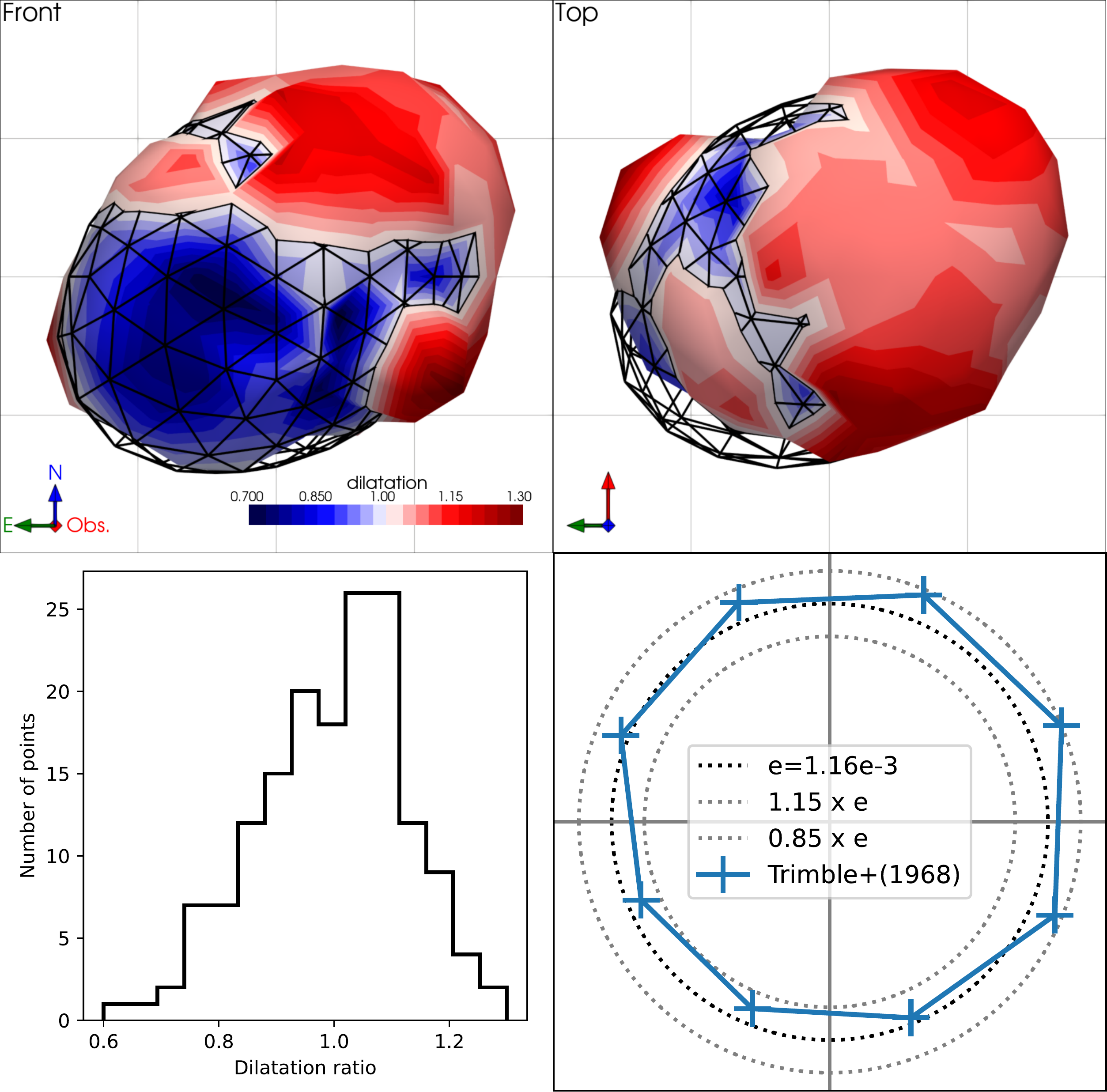}

  \caption{\textit{Top:} Ellipsoid (wireframe) fitted to the outer
  envelope. The ellipsoid represents the surface of dilatation factor unity. Its center coincides with the center of expansion and its major
axis follows the axis of the pulsar torus. The color of the outer envelope represents its dilatation factor. Regions above the ellipsoid are red while regions below are blue. All axes are in
    parsecs and the spatial grid has a 1 pc stepping. The orientation symbol is color coded: East is green, North is blue and the line-of-sight direction is red.\textit{Bottom-left:} Histogram of the dilatation of the
  outer envelope with respect to the fitted
  ellipsoid. \textit{Bottom-right:} median of the expansion factor
  measured by \citet{Trimble1968} along different directions (north is
  up and east to the left). The circles of homologous expansion are
  represented with dotted lines. The north-west acceleration appears
  clearly and can be related to the dilatation along the north-west lobe but there is no acceleration as high as the 1.3 dilatation factor observed in the
  SE-Lobe. Even if we consider the largest expansion factor measured in the plane of the sky (1.15), this is not sufficient to explain the factor 1.3 dilatation observed in the south-west lobe.}

  \label{fig:inhomo}
\end{figure}

The potential 
effects of inhomogeneities of the expansion factor must
be considered since this spatial reconstruction is mostly based on a
homologous expansion hypothesis (i.e., a constant expansion factor in all
directions) which is not absolutely true. Using the data obtained
by \citet{Trimble1968}, we find that the expansion is indeed
accelerated towards the north-west direction by a factor of 15\,\%
with respect to the median value of the expansion factor
($e=1.160(15)\times10^{-3}$\,year$^{-1}$) computed
by \citet{Nugent1998} and used in this article to derive the 3D model
of the nebula (see Figure~\ref{fig:inhomo}). Of course, this
accelerated expansion is associated with a dilatation of the
ellipsoidal shape of the nebula in the same direction. Knowing only
the proper motion of the nebula and using a homologous expansion
model to compute its shape, one would have exaggerated this dilatation
effect, resulting in a nebula with an envelope even more extended
towards the north-east direction. We have fitted an ellipsoid to the
computed outer envelope and obtained the dilatation ratio. The center
of the ellipsoid coincides with the center of expansion and its major
axis follows the axis of the pulsar torus. As shown in
Figure~\ref{fig:inhomo}, the dilatation ratio of the SE-Lobe (which
gives its heart shape to the nebula) is around 1.3. It would require an
expansion inhomogeneity of the same order to keep the model compatible
with an ellipsoid, which is two times larger than what is observed in
the celestial plane. Moreover, this expansion inhomogeneity should not
be related with any extension of the nebula in this direction, which is
in contradiction with the correlation we observed based
on \citet{Trimble1968} data. We are thus confident that, if the
computed 3D model of the nebula might indeed be exaggerated along the
line-of-sight, the general shape should not differ enough to make it
completely compatible with an ellipsoid.

\subsubsection{Inner envelope}
\label{sec:inner_envelope}
The Crab shows an intricate complex of filamentary structures
going deep under its outer envelope, and it can be difficult to easily identify 3D locations of material. Thus, it is of interest to define the inner
extent of this material, i.e. the size and the morphology of the
central void of ionized gas around the center of expansion, to distinguish front-facing from rear-facing ejecta. If we look
at the distribution of material and integrate over all directions
(Figure~\ref{fig:cone_example}) we see that no material is located
below 0.5\,pc. However, we take this a step further by examining the 3D
extent of this void in all directions. Using the same procedure as the
one used to obtain the outer envelope, we obtained the inner envelope
considering a radius enclosing only 8\% of the material emitting
in \Halpha{} in a solid angle of 45\degrees. The limit of 8\,\% may
seem high but is explained by the relatively high number of spurious
detections near the center. We have thus slowly increased the limit up to the
point where the 3D shape of the inner surface was not changing anymore
(except for its scale). Two perspectives of this inner limit surface are shown in
Figure~\ref{fig:innerouter}, and all six perspectives are shown in
Figure~\ref{fig:inner6}.

\subsubsection{Comparison of the inner and outer envelopes}

Comparing the inner and outer envelopes reveals clear trends in the
overall morphology of gas. Material around the plane defined by the
pulsar torus mapped by \citet{Ng2004} is much closer to the explosion
center than the material distributed along the pulsar axis. This can
be seen as a circular pinched valley running along the pulsar torus
plane (labeled {\it equatorial valley}). On the front of the nebula
two small depressions are seen (labeled DB1 and DB2 on the figure)
that coincide with the pinched velocity regions observed by
\citet{MacAlpine89} and the helium-rich bands observed by
\citet{Uomoto1987}. Another indentation is seen in the general region
of the ``arcade of loops'' described by \citet{Dubner2017}.  The two
depressions DB1 and DB2 also coincide with the positions of the dark
bays, \citep{Fesen1992}, also observed in the X-ray \citep{Seward2006}, and UV
(see e.g., \citealt{Dubner2017}). Interestingly, the locations of the dark bays also coincide with a conspicuous restriction oriented along the
east-west plane running along the perimeter of the inner envelope
(labeled {\it inner ring}). Together, these shared features between
the inner and outer envelopes strongly suggest a constrained expansion
of the nebular material, potentially due to interaction with a
pre-existing circumstellar disk left by the progenitor star
\citep{Fesen1992,Smith13}.

\begin{figure*}
  \includegraphics[width=.85\linewidth]{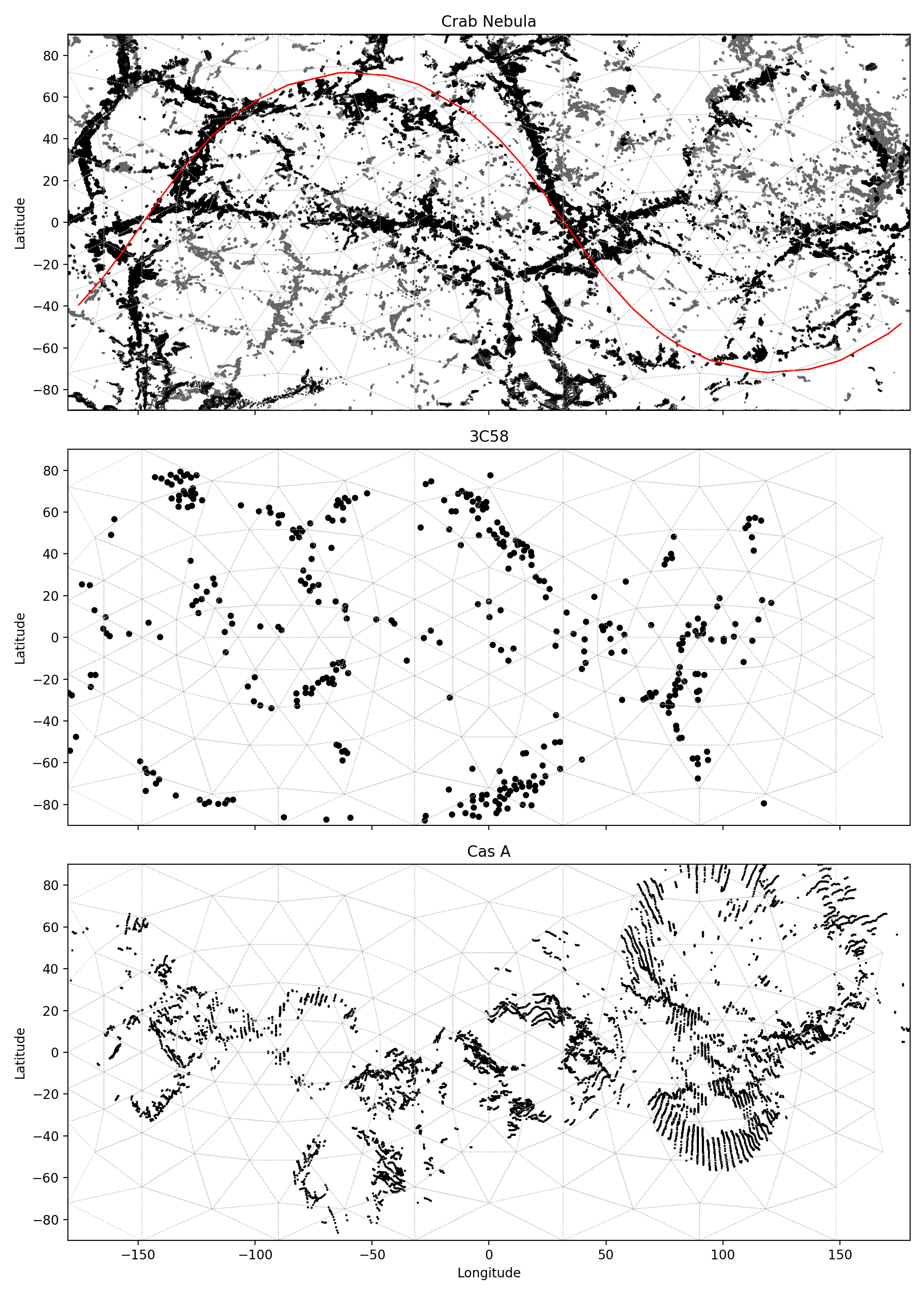}

  \caption{{\it Top}: Mercator projection of total flux of the
    emitting material in the Crab. The densest and deepest buried
    material (at less than 1.1\,pc of the center) is represented in
    black while the faster expanding material is in gray. As the angles
    are not conserved, we overplot a grid of triangles which vertices
    sides cover an angle of 16.6\degrees ($\simeq\pi/11$). The red line shows the
    pulsar torus plane as fitted by \citet{Ng2004}. The center of the latitude and longitude axes corresponds to the
center of expansion in the face-on view. Longitude increases towards the West. {\it Middle}
    and {\it bottom}: Mercator projections of 3C\,58 \citep{Lopez2018}, 
    and Cas A \citep{Mili13}, respectively.}
  \label{fig:mercat_comparison}  
\end{figure*}

\subsection{Filamentary structure}
\label{sec:filamentary_structure}

High resolution images of the Crab show that its filaments exhibit a
complex and fine structure (see, e.g., \citealt{Hester1996},
\citealt{Blair97}, and \citealt{Sankrit98}). Many filaments are less
than an arcsecond in width and point inward into the center of the
nebula, with lengths ranging from
$\approx 1^{\prime\prime} - 20^{\prime\prime}$. Filaments are also
often connected by arc-like bridges of emission with a
``bubble-and-spike'' morphology \citep{Hester2008}. Numerous studies
have associated this morphology with Rayleigh-Taylor
(RT) instabilities (see for example
\citealt{CG75,Hester1996,Bucciantini2004,Stone2007,Porth2014}).  These
instabilities are generally characterized by 2 parameters. (1) Angular
size, i.e., the wavelength of the perturbations, which is strongly
related to the stability of the shell. Numerical MHD simulations
\citep{Bucciantini2004} show that only perturbations at a scale
smaller than $\sim\pi/10$ should give rise to RT instabilities and to
the observed structures with protruding fingers. (2) The size of the
filaments, which should also be related to the wavelength of these
perturbations.

\begin{figure*}
  \includegraphics[width=1\linewidth]{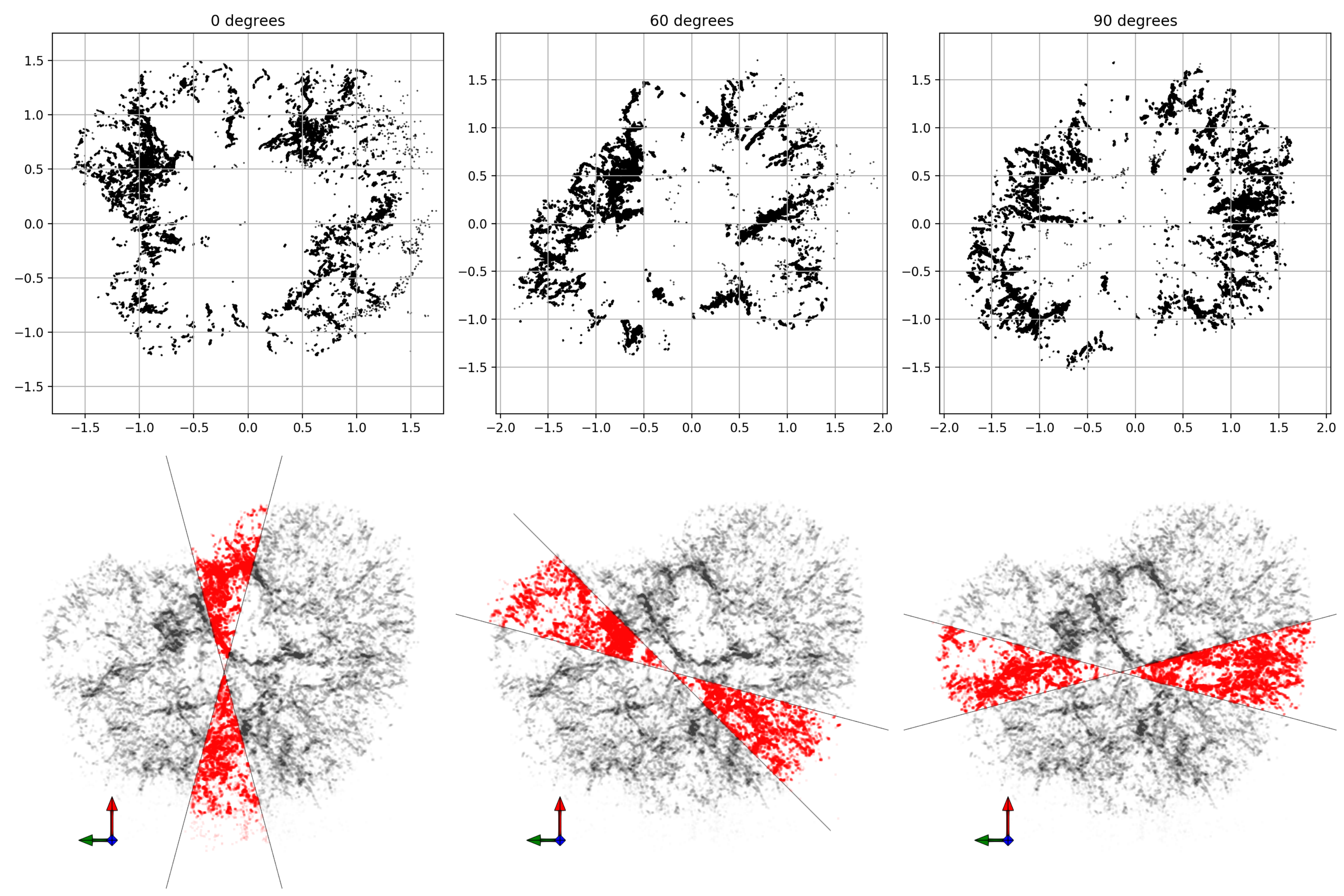}
  \caption{Slices in 3 different directions with respect to the line
    of sight direction. The slice at 90\degrees{} beeing the canonical
    point of view perpendicular to the line of sight. All points at
    $\pm$15\degrees are merged together so that the angular size of a
    slice is 30\degrees. A top view of the selected pixels is shown at
    the bottom. It shows the slice extent
used to generate the plots in the corresponding upper panel. Axes are in pc. The orientation symbol is color coded: East is green, North is blue and the line-of-sight direction is red.}
  \label{fig:slices_3}
\end{figure*}

In Figure~\ref{fig:mercat_comparison} we show the whole structure as
if all voxels were at the same radius in a classical Mercator
projection, and in Figure~\ref{fig:mercat} as orthographic
projections. All maps are gridded with triangles having sides covering
an angle of 16.6\degrees ($\simeq\pi/11$), which is helpful to
measure the relative sizes of the different structures.

Crab material is largely distributed along boundaries resembling a
honeycomb. This structure is hierarchical; i.e., larger regions
($\sim\pi/3.5$) have thicker filaments, while other regions, in
particular in the high-velocity lobes, exhibit a much smaller and
thinner distribution ($\sim\pi/10$). The interior of the largest
structures is generally divided into smaller regions that are also at
a larger radii.  The size of the structures appear to be
anti-correlated with the radius; i.e., the largest and most
conspicuous structures are at smallest radii with smallest expansion
velocities. Some of the largest structures, seen clearly on the top
and back views of Figure~\ref{fig:mercat_comparison}, exhibit a
polygonal shape and are found all along the pulsar torus equatorial
plane. One also sees these structures when viewing the Crab through
narrow slices along different orientations
(Figure~\ref{fig:slices_3}). The largest regions are localized in the
interior with diameters of 0.5 to 1\,pc. Above these at larger radii
lie numerous smaller bubbles.

\section{Discussion}
There exist very few kinematic maps of supernova remnants detailed
enough that a comparison of the shape of their filamentary structure
can be studied. Figure~\ref{fig:mercat_comparison} shows a comparison
of the mercator projections of the Crab, 3C\,58 \citep{Lopez2018} and
Cas~A \citep{Mili13}. Large-scale ejecta rings may be a common
phenomenon of young, core-collapse SNRs \citep{MF17}. The largest
and deepest structures of the Crab are similar both in size and shape
as those seen in 3C\,58. However, both the Crab and Cas~A share small
scale circular formations.

3C\,58 has many overlapping properties with the Crab. The Crab Nebula is far
brighter and more luminous than 3C\,58, but both remnants are bright
in both the radio and X-rays in their center and harbor young, rapidly
spinning central pulsars that provide the magnetic field and
relativistic particles that generate the observed center-filled
synchrotron radiation. 3C\,58 may be connected to the
historical event of 1181\,CE \citep{Stephenson2002, Kothes2013},
which would make it only 127 years younger than the Crab. However,
this relatively young age is inconsistent with its overall angular
size ($6.3^{\prime} \times 10.3^{\prime}$) and proper motion
measurements of its expanding ejecta \citep{Fesen88,vandenBergh90}
that suggest a much older remnant, potentially as old as
$2400 \pm 500$\,yr.

Cas~A is the youngest known core-collapse remnant with an estimated
explosion date of 1681\,CE \citep{Fesen2006}. \citet{Mili13}
demonstrated that the bulk of the remnant's main shell
ejecta are arranged in several well-defined complete or broken
ring-like structures. These ring structures have diameters that can be
comparable to the radius of the remnant ($\sim 1$ pc). Some rings show
considerable radial extensions giving them a crown-like appearance,
while other rings exhibit a frothy, ring-like substructure on scales
of $\sim 0.2$ pc.  A subsequent three-dimensional map of its interior
unshocked ejecta made from near-infrared observations sensitive to
[S~III] $\lambda\lambda$ 9069, 9531 emission lines revealed a
bubble-like morphology that smoothly connects with these rings
\citep{MF15}.

In the case of Cas~A, the rings of ejecta have been interested to be
the cross sections of reverse-shock-heated cavities in the remnant's
internal ejecta \citep{Mili13}. A cavity-filled interior is in line
with prior predictions for the arrangement of expanding debris created
by a post-explosion input of energy from plumes of radioactive
$^{56}$Ni-rich ejecta \citep{Li93,Blondin01}. Such plumes can push the
nuclear burning zones located around the Fe core outward, creating
dense shells separating zones rich in O, S, and Si from the Ni-rich
material. After the SN shock breakout additional energy input from the
radioactive decay of $^{56}$Ni continues to drive inflation of $^{56}$Ni-rich
structures and facilitates mixing between ejecta components. This late
time expansion can modify the overall SN ejecta morphology on
timescales of weeks or months. Compression of surrounding
nonradioactive material by hot expanding plumes of radioactive
$^{56}$Ni-rich ejecta generates a ``Swiss cheese''-like structure that is
frozen into the homologous expansion when the radio- active power of
$^{56}$Ni is strongest. \citet{Gabler20} found that this ``Ni-bubble
effect'' accelerates the bulk of the nickel in their 3D models and
causes an inflation of the initially overdense Ni-rich clumps, which
leads to underdense, extended fingers, enveloped by overdense skins of
compressed surrounding matter.

\citet{Stockinger20} recently performed 3D full-sphere simulations of
supernovae originating from non-rotating progenitors similar to those
anticipated to be associated with the SN\,1054. Their low energy
explosions ($\sim 0.5 - 1.0 \times 10^{50}$\,erg) are compared in two
contrasting scenarios: (1) iron-core progenitors at the low-mass end
of the core-collapse supernova domain ($\approx 9$\,M$_{\odot}$), and
(2) a super-AGB progenitor with an oxygen-neon-magnesium core that
collapses and explodes as electron-capture supernova. They disfavor
associating SN\,1054 with an electron capture supernova because the
kick experienced by the neutron star is negligible and inconsistent
with the observed $\approx 160$ \kms\ transverse velocity of Crab
pulsar. Instead, they favour simulations with iron-core progenitors
with less 2nd dredge-up that result in highly asymmetric explosions
with hydrodynamic and neutrino-induced NS kick of $>40$ \kms\ and a NS
spin period of $\sim 30$\,ms, not unlike the Crab pulsar. The
resulting distribution of $^{56}$Ni-rich material from these
explosions, which enable efficient mixing and dramatic shock
deceleration in the extended hydrogen envelope, is potentially
consistent with our mapping of Crab ejecta. However, simulations
extending the evolution from $\sim$ days to $\sim$ 1000 years are
needed to verify this extrapolation (see, e.g., \citealt{Orlando16}).

\section{Conclusions}

We have presented a 3D kinematic reconstruction of the Crab Nebula
that has been created from a hyperspectral cube obtained with
SITELLE. The data is comprised of 310\,000 high resolution
($R = 9\,600$) spectra containing \Halpha, [\NII], and [\SII] line
emission, and represent the most detailed homogeneous spectral data
set ever obtained of the Crab Nebula.

Our findings can be summarized as follows:

\begin{enumerate}

\item The general shape of the Crab, as measured by 97\% of the
  material emitting in \Halpha, occupies a ``heart-shaped'' volume and
  is symmetrical about the plane of the pulsar wind torus. This
  morphology runs counter to the generally assumed ellipsoidal
  volume and is not an artifact of assuming a uniform global expansion. The most rapidly expanding NW and SE lobes are separated by
  120\degrees{} of each other. The NW lobe is nearly aligned with the
  pulsar torus axis, but the SE lobe is not. 

\item Conspicuous restrictions in the distribution of material as
  mapped by the inner and outer limits of emission is seen along the
  band of He-rich filaments \citep{Uomoto1987,MacAlpine1989}. Notable
  depressions are also coincident with the east and west dark bays
  \citep{Fesen1992}. Together these features are consistent with
  constrained expansion of Crab ejecta, possibly associated with
  interaction between the supernova and a pre-existing cirumstellar
  disk.

\item The filaments follow a honeycomb-like distribution defined by a
  combination of straight and rounded boundaries at large and small
  scales. The scale size is anti-correlated with distance from the
  center of expansion; i.e., largest features are found at smallest
  radii. The structures are not unlike those seen in other SNRs,
  including 3C\,58 and Cas~A, where they have been attributed to
  turbulent mixing processes that encouraged outwardly expanding
  plumes of radioactive $^{56}$Ni-rich ejecta.

\end{enumerate}

The observed kinematic characteristics reflect critical details
concerning the original supernova of 1054\,CE and its progenitor star,
and may favour a low-energy explosion of an iron-core progenitor as
opposed to an oxygen-neon-magnesium core that collapses and explodes
as an electron-capture supernova. Planned future observations will
provide additional hyperspectral cubes spanning more wavelength
windows that include the emission lines of [O~II]
$\lambda\lambda$3726, 3729, H$\beta$, [O~III] $\lambda\lambda$4959,
5007, [N~II] $\lambda$5755, and He~I $\lambda$5876. These lines will
be measured and modeled to determine temperature, density and
abundances at very fine scales, and combined with an updated proper
motion investigation of filaments (Martin et al., in preparation) to
improve the accuracy of our reconstruction at fine
scales. Our work contributes to a larger suite of
detailed supernova remnant reconstructions being developed that will
provide unique constraints for increasingly sophisticated
three-dimensional core-collapse simulations
\citep{Couch15,Wong17,Burrows19,Stockinger20}, that are being evolved
to middle-aged supernova remnants
\citep{Orlando15,Orlando16,Orlando20}, attempting to model the
complete multi-messenger signals of supernovae
\citep{Andresen17,Kuroda17,West19,Mezz20}.

\section*{Data availability}
The data underlying this article will be shared on reasonable request to the corresponding author.

\section*{Acknowledgements}

We thank the referee for comments that greatly improved the paper. We thank R.\ Fesen for many helpful discussions that guided
interpretation of our data, and Hans-Thomas Janka who commented on an earlier draft of the manuscript. We are also thankful to the
\texttt{python} \citep{VanRossum2009} community and the free
softwares that made the analysis of this data possible: \texttt{numpy}
\citep{Oliphant2006}, \texttt{scipy} \citep{Virtanen2020},
\texttt{pandas} \citep{McKinney2010}, \texttt{panda3d}
\citep{Goslin2004}, \texttt{pyvista} \citep{Sullivan2019},
\texttt{matplotlib} \citep{Hunter2007} and \texttt{astropy}
\citep{Price2018}. This paper is based on observations obtained with
SITELLE, a joint project of Universit{\'e} Laval, ABB, Universit{\'e}
de Montr{\'e}al and the Canada-France-Hawaii Telescope (CFHT) which is
operated by the National Research Council (NRC) of Canada, the
Institut National des Science de l'Univers of the Centre National de
la Recherche Scientifique (CNRS) of France, and the University of
Hawaii.  The authors wish to recognize and acknowledge the very
significant cultural role that the summit of Mauna Kea has always had
within the indigenous Hawaiian community.  LD is grateful to the
Natural Sciences and Engineering Research Council of Canada, the Fonds
de Recherche du Qu{\'e}bec, and the Canadian Foundation for Innovation
for funding. DM acknowledges support from the National Science
Foundation from grants PHY-1914448 and AST-2037297.





\bibliographystyle{mnras}
\bibliography{m1} 




\appendix
\section{Velocity Uncertainty}
\label{sec:uncertainty}
\begin{figure*}
  \includegraphics[width=.9\linewidth]{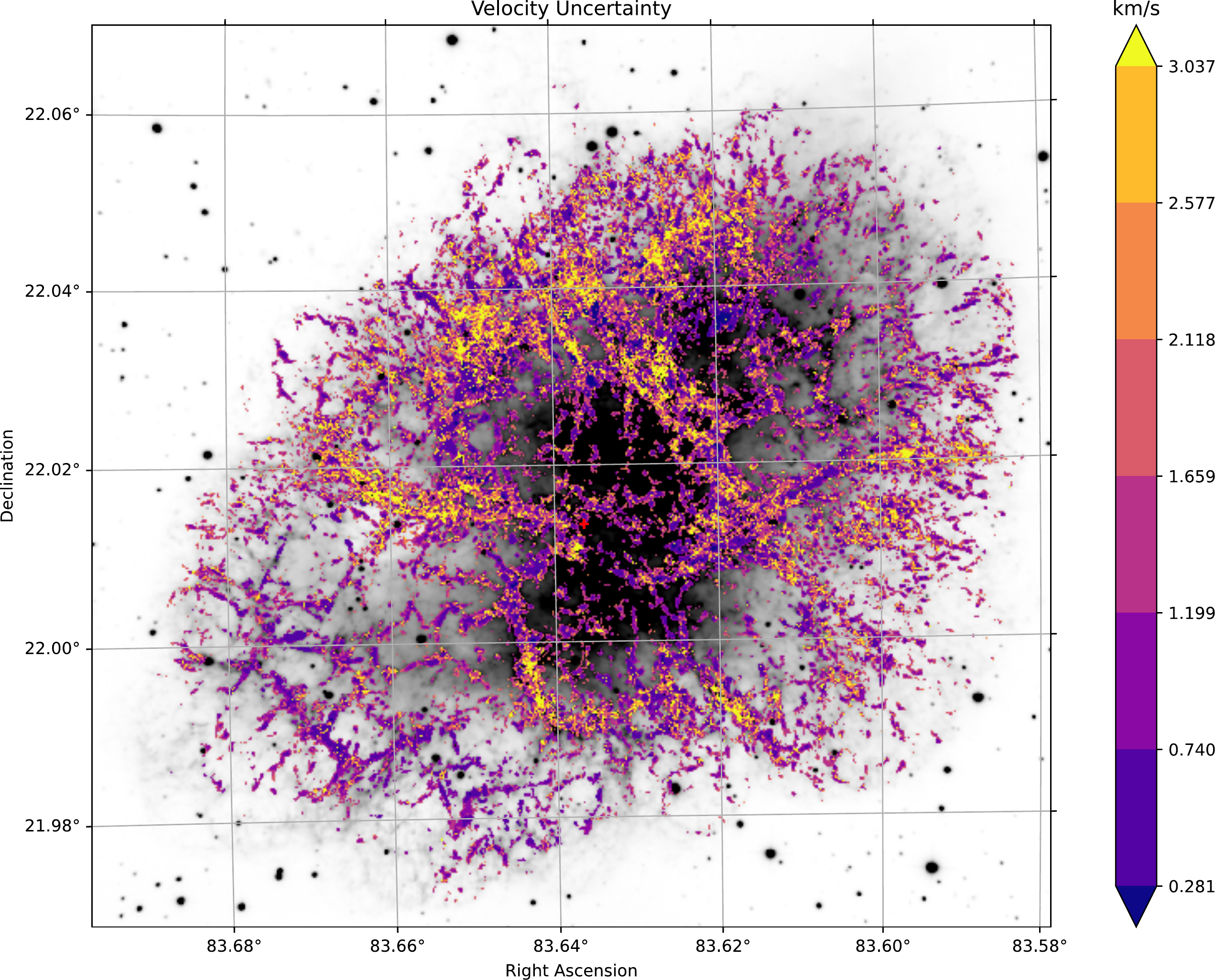}
  \caption{Map of the velocity uncertainty related to Figure~\ref{fig:fig_all}. The uncertainty is an output of the fit realised with \texttt{ORCS} \citep{Martin2015}. It is calculated from the covariance matrix returned by the Levenberg-Marquardt minimization process \citep{Levenberg1944, Marquardt1963}. As all 5 emission-lines for each velocity component share the same velocity parameter, the uncertainty is generally smaller that the uncertainty that would have been obtained by fitting the emission-lines independently. As other constraints are implemented in the spectrum model (see section~\ref{sec:step3}) the emission-lines velocities and fluxes are the parameters of one spectrum model. As such their uncertainty are only loosely related to the SNR of each emission-line (see \citealt{Martin2015} for more details).}
  \label{fig:fig_all_err}
\end{figure*}

\section{3d maps}
\label{sec:3dmaps}

\begin{figure*}
  \includegraphics[width=.8\linewidth]{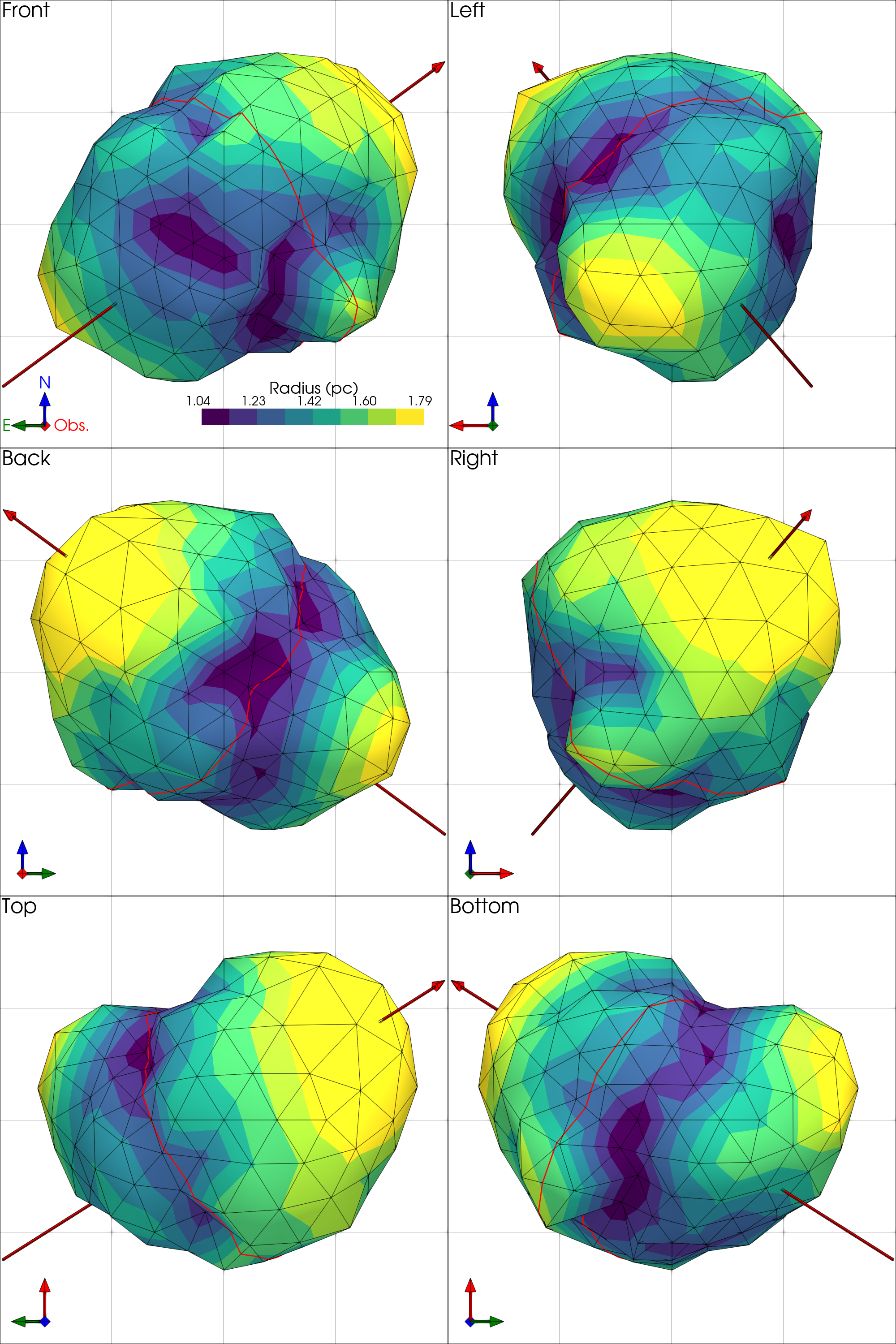}
  \caption{Outer envelope related to Figure~\ref{fig:innerouter}. All 6
    viewing angles are shown. Details are the same as
    Figure~\ref{fig:all_3d} and \ref{fig:innerouter}. The orientation symbol is color coded: East is green, North is blue and the line-of-sight direction is red.}
  \label{fig:outer6}
\end{figure*}

\begin{figure*}
  \includegraphics[width=.8\linewidth]{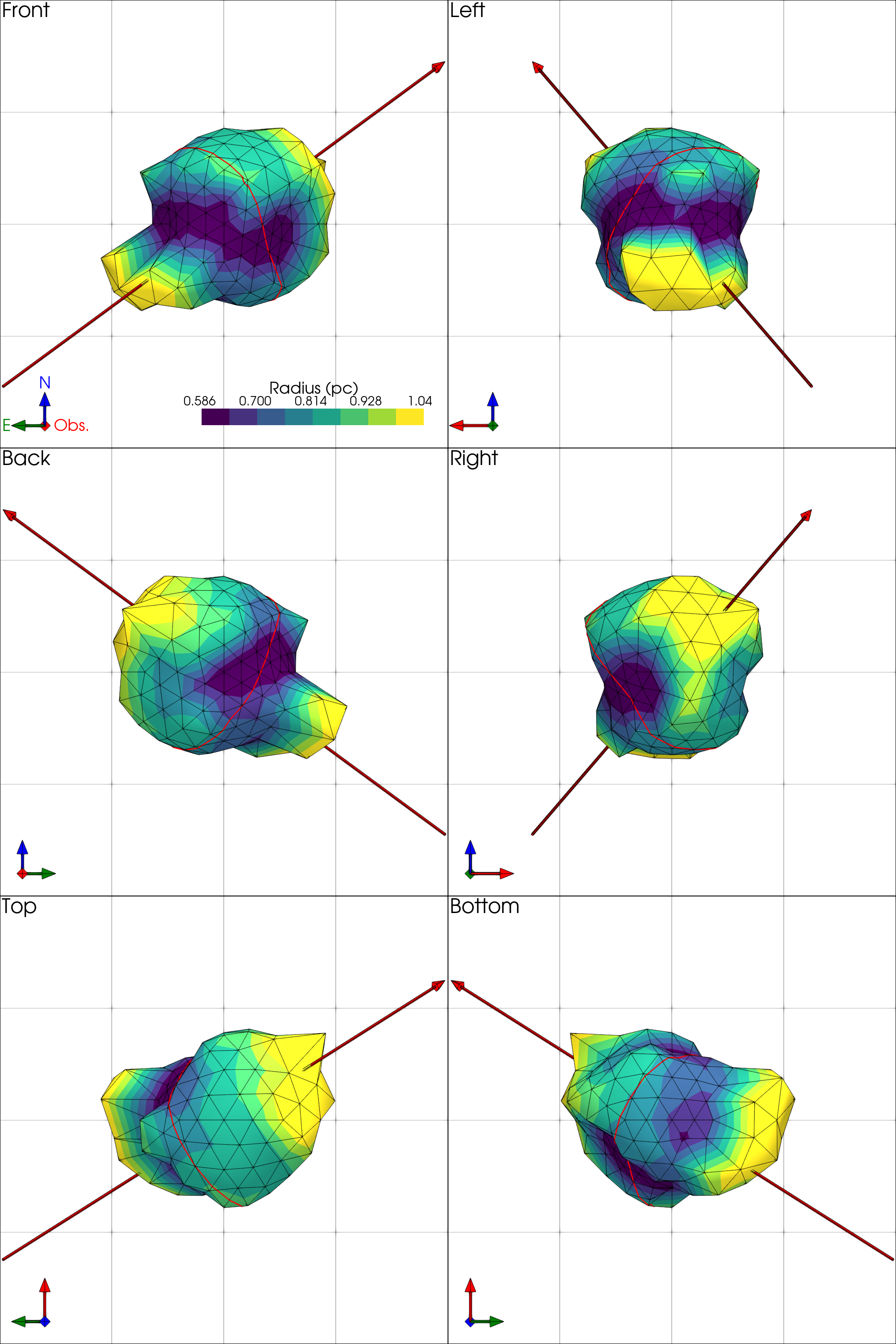}
  \caption{Inner envelope related to Figure~\ref{fig:innerouter}. All
    6 viewing angles are shown. Details are the same as
    Figure~\ref{fig:all_3d} and \ref{fig:innerouter}. The orientation symbol is color coded: East is green, North is blue and the line-of-sight direction is red.}
  \label{fig:inner6}
\end{figure*}

\section{Icosphere}
\begin{figure*}
  \includegraphics[width=.8\linewidth]{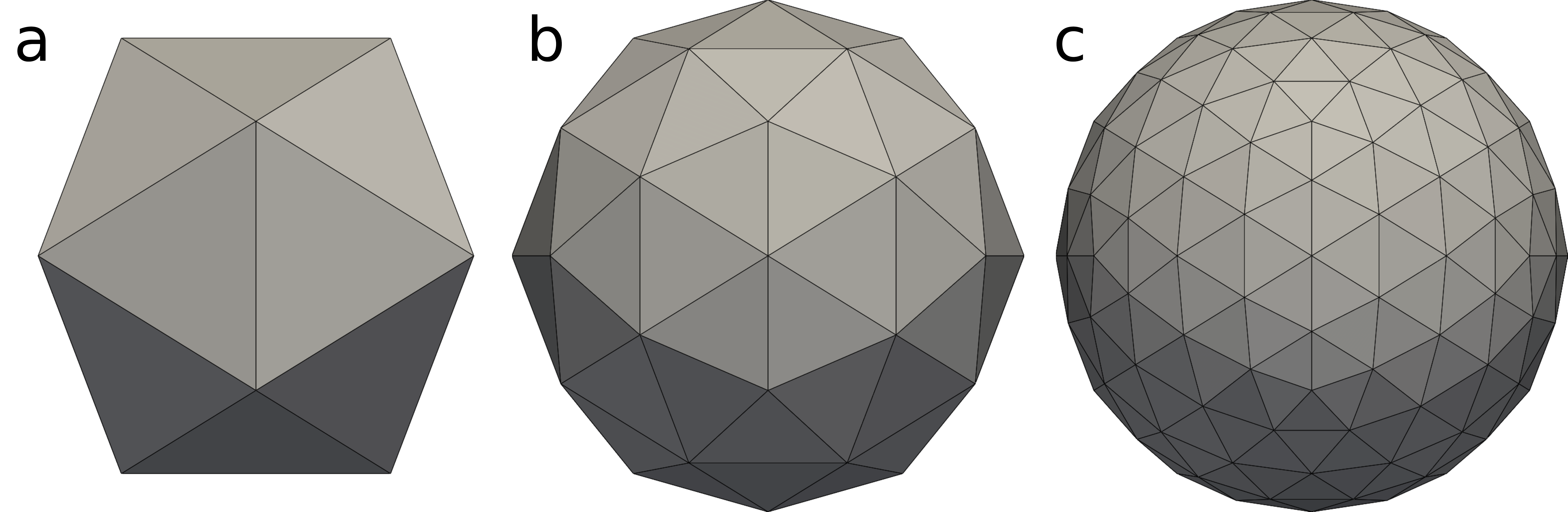}
  \caption{Icosahedron (\textit{a}, 20 faces), subdivided 2 times
    along its edges to make the 80 (\textit{b}) and 320 faces
    (\textit{c}) icospheres. This type of sphere approximation is
    interesting since its vertices are homogeneously distributed. We
    used the 320 icosphere to construct the inner and outer envelope
    as well as to make a homogeneous grid on the projections of
    Figures~\ref{fig:mercat_comparison} and~\ref{fig:mercat}.}
    \label{fig:icosphere}
\end{figure*}

\section{Filamentary structure}

\begin{figure*}
  \includegraphics[width=.8\linewidth]{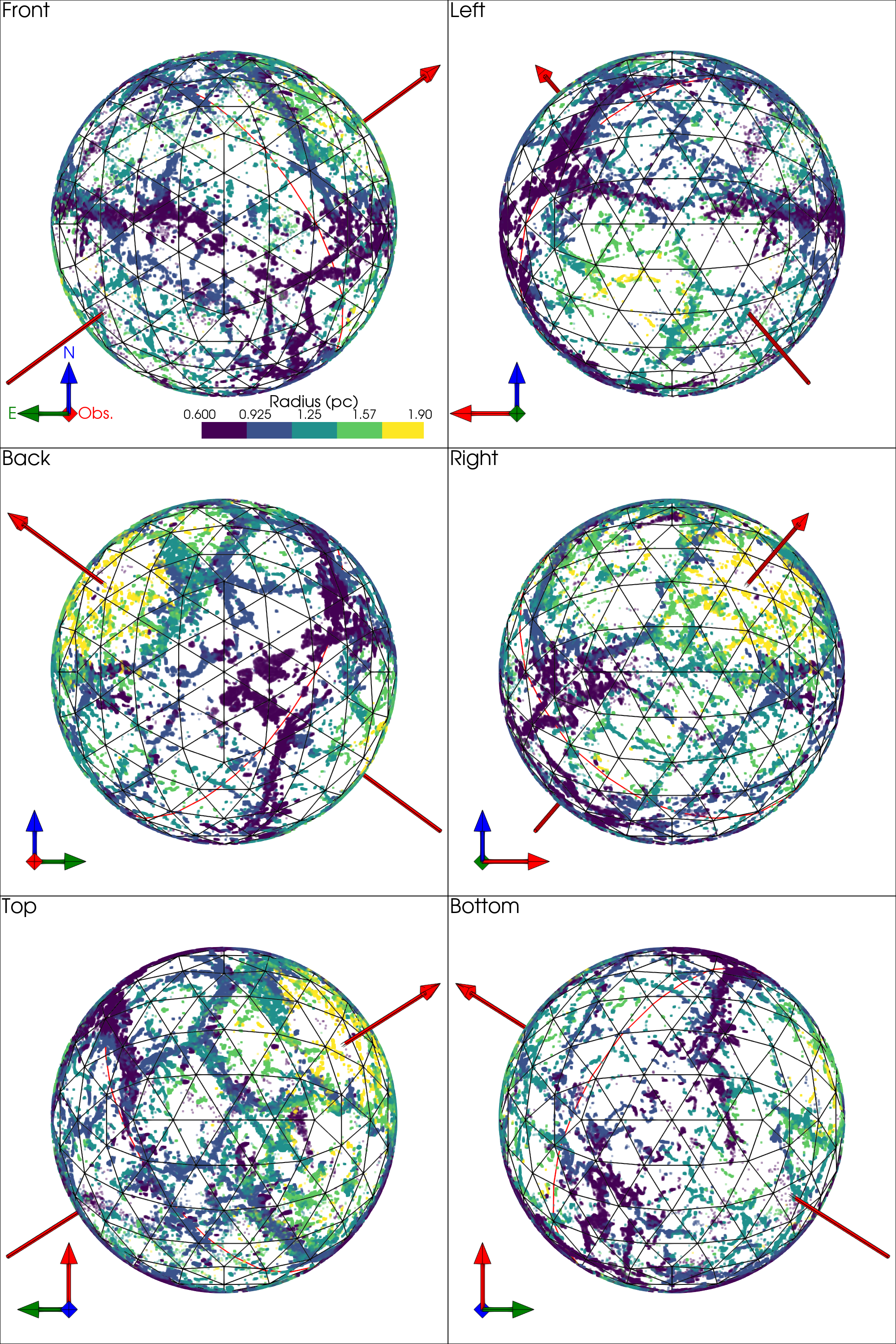}
  \caption{Orthographic projections of all the voxels at the same unit radius to
    reveal the structuration of the filamentary envelope of the
    nebula. This is presented as an alternative to Mercator projections shown in Figure~\ref{fig:mercat_comparison} which produce important distortions far from the equator. A grid of triangles with 16.6\degrees ($\simeq\pi/11$)
    sides helps to measure the size of the structures. The red arrow
    indicates the direction of the pulsar torus and the red line shows
    its equator. The orientation symbol is color coded: East is green, North is blue and the line-of-sight direction is red.}
  \label{fig:mercat}  
\end{figure*}


\bsp	
\label{lastpage}
\end{document}